\documentclass[10pt,twocolumn,letterpaper]{article}

\usepackage{wacv}
\usepackage{times}
\usepackage{epsfig}
\usepackage{graphicx}
\usepackage{amsmath}
\usepackage{amssymb}
\usepackage{tabularx}
\usepackage{adjustbox}
\usepackage{float}
\usepackage{graphicx}
\usepackage[utf8]{inputenc}
\usepackage{newunicodechar}
\usepackage{multirow}
\usepackage{caption}
\usepackage{balance}
\usepackage{array}
\usepackage{subfig}
\usepackage{xcolor}
\usepackage{soul}

\usepackage{booktabs}
\newunicodechar{ℝ}{\mathbb{R}}
\usepackage{tabularx}
\usepackage{pifont}% http://ctan.org/pkg/pifont
\newcommand{\cmark}{\ding{51}}%
\newcommand{\xmark}{\ding{55}}%

\usepackage[sort,square,numbers]{natbib}
\usepackage[font=small,skip=4pt]{caption}

\usepackage{color}
\definecolor{red}{rgb}{1.00,0.00,0.00}
\definecolor{blue}{rgb}{0.00,0.00,1.00}
\definecolor{green}{rgb}{0.4,1.00,0.0}
\definecolor{yellow}{rgb}{0.5,0.5,0.0}

\def\wacvPaperID{953} % Enter the WACV Paper ID here

\wacvfinalcopy % *** Uncomment this line for the final submission
\ifwacvfinal
\def\assignedStartPage{9876} % *** Enter the assigned starting page number (instead of 9876)
\fi

\ifwacvfinal
\usepackage[breaklinks=true,bookmarks=false]{hyperref}
\else
\usepackage[pagebackref=true,breaklinks=true,colorlinks,bookmarks=false]{hyperref}
\fi

\ifwacvfinal
\else
\fi

\begin{document}

%%%%%%%%% TITLE
\title{MPRNet: Multi-Path Residual Network for Lightweight Image Super Resolution}
% \title{HLNet: Hyper Lightweight Image Super Resolution with Multi-Path Residual Block}

\author{Armin Mehri \\
$^{*}$Computer Vision Center,\\
Edifici O, Campus UAB, \\
08193, Bellaterra,\\
Barcelona, Spain\\
{\tt\small amehri@cvc.uab.es}
\and
Parichehr B.Ardakani \\
$^{*}$Computer Vision Center,\\
Edifici O, Campus UAB, \\
08193, Bellaterra,\\
Barcelona, Spain\\
{\tt\small pbehjati@cvc.uab.es}
\and
Angel D. Sappa \\
$^{+}$ESPOL Polytechnic University,\\
Guayaquil, Ecuador \vspace{5pt} \\
$^{*}$Computer Vision Center,\\
08193, Bellaterra, Barcelona, Spain\\
{\tt\small sappa@ieee.org}}

% \author{First Author\\
% Institution1\\
% Institution1 address\\
% {\tt\small firstauthor@i1.org}
% % For a paper whose authors are all at the same institution,
% % omit the following lines up until the closing ``}''.
% % Additional authors and addresses can be added with ``\and'',
% % just like the second author.
% % To save space, use either the email address or home page, not both
% \and
% Second Author\\
% Institution2\\
% First line of institution2 address\\
% {\tt\small secondauthor@i2.org}
% }

\maketitle
%\thispagestyle{empty}
% The proposed architecture, make the network to pay attention on learning more abstract features by letting abundant low-frequency features to be avoided via multiple connections.

% % Lightweight super resolution networks have extremely
% importance for real-world applications. Recent years
% several SR deep learning approaches with outstanding
% achievement have been introduced, however their memory and computational cost are obstacles in practical usage. To overcome these problems, 

    % Deep neural networks with a massive number of layers have
    % made a remarkable breakthrough on single image super-resolution (SR), but sacrifice computation complexity and memory storage.

%%%%%%%%% ABSTRACT
\begin{abstract}
    Lightweight super resolution networks have extremely importance for real-world applications. In recent years several SR deep learning approaches with outstanding achievement have been introduced by sacrificing memory and computational cost. To overcome this problem, a novel lightweight super resolution network is proposed, which improves the SOTA performance in lightweight SR and performs roughly similar to computationally expensive networks. Multi-Path Residual Network designs with a set of Residual concatenation Blocks stacked with Adaptive Residual Blocks: ($i$) to adaptively extract informative features and learn more expressive spatial context information; ($ii$) to better leverage multi-level representations before up-sampling stage; and ($iii$) to allow an efficient information and gradient flow within the network. The proposed architecture also contains a new attention mechanism, Two-Fold Attention Module, to maximize the representation ability of the model. Extensive experiments show the superiority of our model against other SOTA SR approaches.
\end{abstract}

%%%%%%%%% BODY TEXT
\vspace{-20.0pt}
\section{Introduction} \label{sec:intro}

Single Image Super Resolution (SISR) targets to recover a high-resolution (HR) image from its degraded low-resolution (LR) one with a high visual quality and enhanced details. SISR is still an active yet challenging topic to research due to its complex nature and high practical values in improving image details and textures. SR is also critical for many devices such as HD TVs, computer displays and portable devices like cameras, smartphones, tablets, just to mention a few. Moreover, it leads to improvements in various computer vision tasks, such as object detection \cite{girshick2015region}, medical imaging \cite{greenspan2008super}, security and surveillance imaging \cite{zou2011very}, face recognition \cite{mudunuri2015low}, astronomical images \cite{lobanov2005resolution} and many other domains \cite{wang2016studying,yu2018generative,liu2017robust}. Image super-resolution is challenging due to the following reasons: $i)$ SR is an ill-posed inverse problem, since instead of a single unique solution, there exist multiple solutions for the same low-resolution image; and $ii)$ as the up-scaling factor increases, the complexity of the problem increases \cite{dong2014learning}. The retrieval of missing scene details becomes even more complicated with greater factors, which often leads to the reproduction of incorrect information. 
% $iii)$, evaluation of the quality of output is not straightforward i.e., quantitative methttps://sgsrevisionestecnicas.ec/rics (e.g. PSNR, SSIM \cite{wang2004image}) only loosely correlate to human perception \cite{anwar2019deep}.
\begin{figure}[t!]
        \centering
        \includegraphics[width=1.0\linewidth]{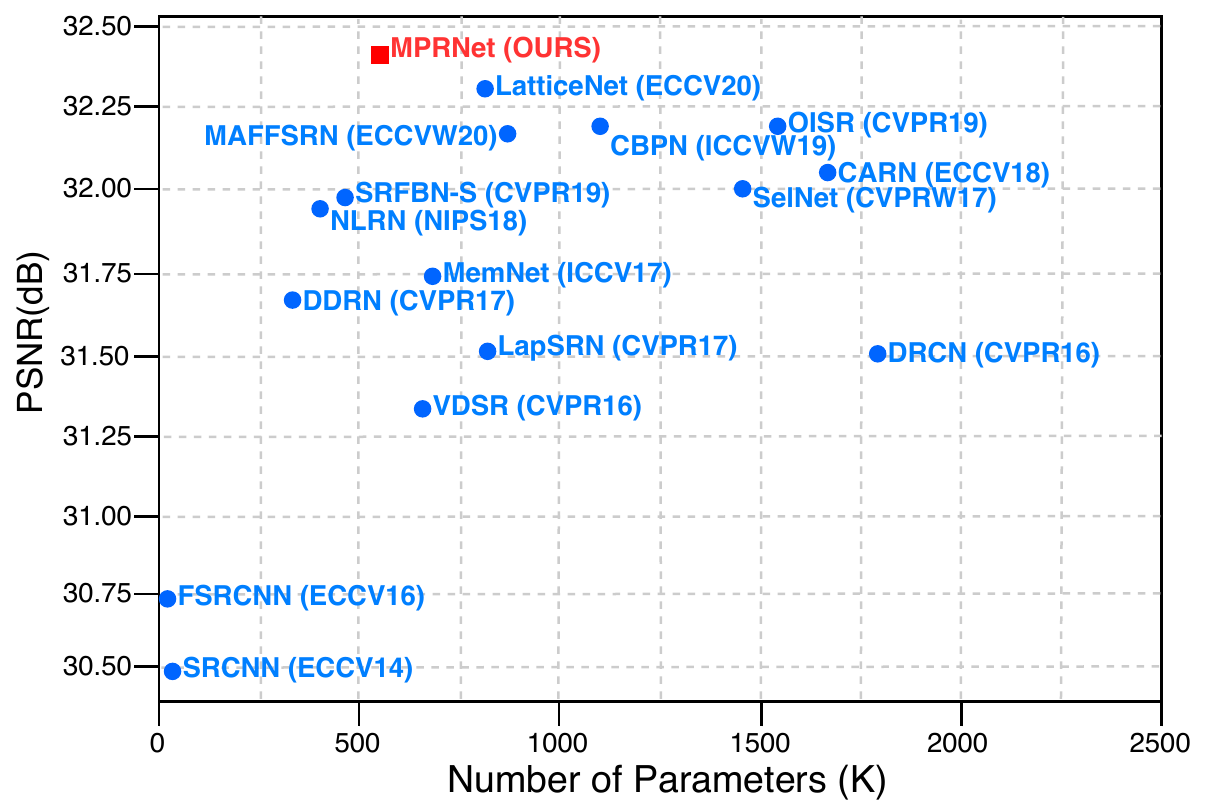}
        % \vspace{5pt}
        \caption{\small PSNR $vs.$ Parameters trade-off on Set$5$ ($\times4$). MPRNet achieves superior performance among all lightweight models.}
        \label{fig:ParamsMac}
\vspace{-0.3cm}
\end{figure}
Due to the rapid development of deep learning methods, recent years have witnessed an explosive spread of CNN models to perform SISR. The obtained performance has been consistently improved by designing new architectures or introducing new loss functions. Though significant advances have been made, most of the works in SR were dedicated to achieve higher PSNR with the design of a very deep network, which causes the increase in the numbers of computational operations. 
% Besides that, most of the existing SISR methods are trained and evaluated on simulated datasets that assume simple and bicubic degradation. Unfortunately, SISR models trained on such simple datasets are hard to generalize for practical applications since degradations in the real-world are unknown.

\begin{figure*}
        \centering
        \includegraphics[width=\textwidth]{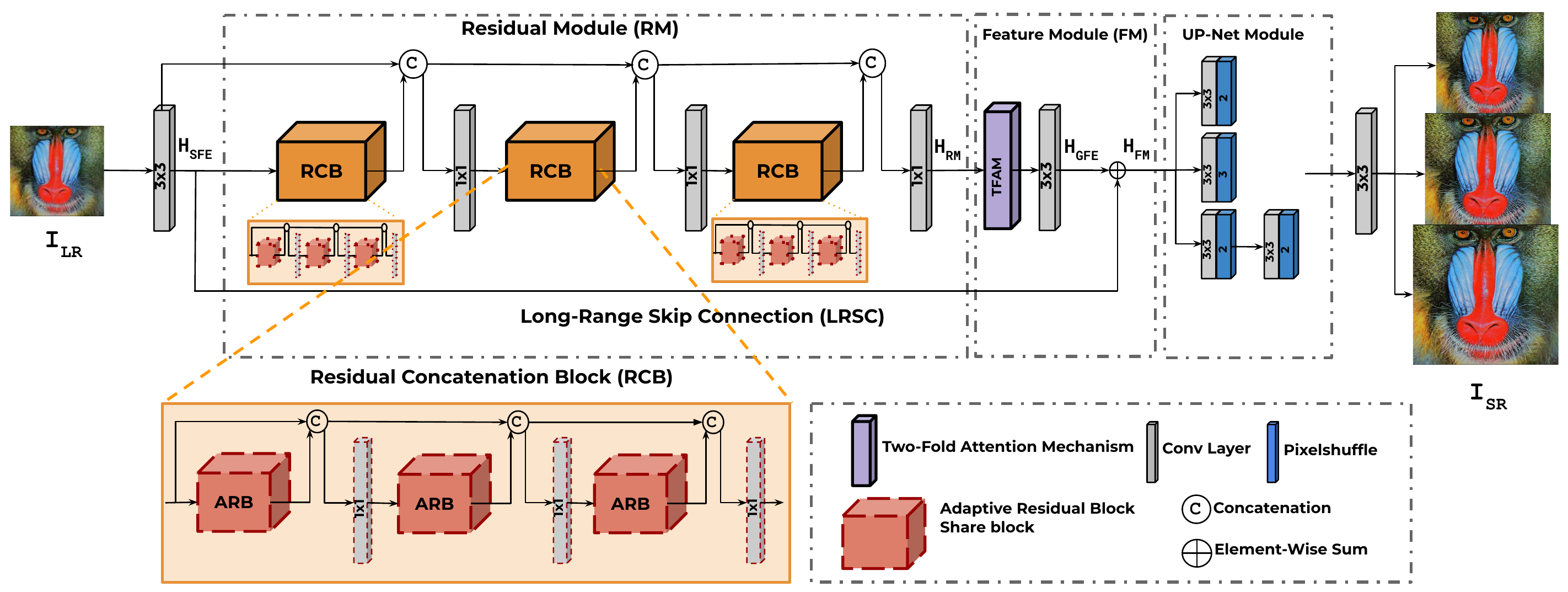}
        \caption{The overall network architecture of the proposed Multi-Path Residual Network (MPRNet).}
        \label{fig:arc}
        % \vspace{-10pt}
\end{figure*}

In this paper, to design a practical network for real-world applications and tackle with mentioned downsides, a novel lightweight architecture is introduced, referred to as Multi-Path Residual Network (MPRNet), to adaptively learn most valuable features and construct the network to focus on learning high-frequency information. Additionally, to seek a better trade-off between performance and applicability, we introduce a novel module, referred to as Residual Module (RM), which contains Residual Concatenation Blocks that are connected to each other with a Global Residual connection; build with a set of Adaptive Residual Blocks (ARB) with a Local Residual Connection (LRC). Each ARB is defined as a divers residual pathways learning to make use of all kind of information form LR image space, which the main parts of the network can access to more rich information. So, our MPRNet design has the benefits of multi-level learning connections and also takes advantage of propagating information throughout the network. As a result, each block has access to information of the precedent block via local and global residual connections and passes on information that needs to be preserved. By concatenating different blocks followed by $1\times1$ convolutional layer the network can reach to both intermediate and high-frequency information, resulting in a better image reconstruction. Finally, in order to enhance the representation of the model and even make it robust against challenging datasets and noise, we propose a lightweight and efficient attention mechanism, Two-Fold Attention Mechanism (TFAM). TFAM is working by considering both the inner channel and spatial information to highlight the important information. This TFAM helps to adaptively preserve essential information and overpower the useless ones. The proposed model is illustrated in Fig. \ref{fig:arc}. In brief, the main contributions are in three-fold:

\begin{itemize}

\item An efficient Adaptive Residual Block (ARB) is proposed by well-focusing on spatial information via a multi-path residual learning to enhance the performance at a negligible computational cost. Comprehensive study shows the excellent performance of ARB.

\item A new attention mechanism (TFAM) is proposed to adaptively re-scale feature maps in order to maximize the representation power of the network. Since its low-cost, it can be easily applied to other networks, and has the better performance than other Attention Mechanisms.

\item A lightweight network (MPRNet) is proposed to effectively enhance the performance via multi-level representation and multiple learning connections. The MPRNet is built by fusing the proposed ARB with the robust TFAM to generate more accurate SR image. MPRNet achieves the excellent performance among all the lightweight state-of-the-art approaches with lower model size and computational cost (Fig. \ref{fig:ParamsMac}).

% MPRNet achieves the superior performance on several SR benchmark
% datasets against the state-of-the-art methods while maintaining relatively
% lower model size and computational complexity.

% \vspace{-12pt}
\end{itemize}
%------------------------------------------------------------------, section \ref{sec:dl_sr}

\vspace{-0.5cm}
\section{Related Work} \label{sec:related}

In this section, recent state-of-the-art SR deep learning approaches are detailed. In section \ref{sec:dl_wsr}, SR lightweight models, which focus on compressing the number of parameters and operations are reviewed. Finally, an overview of Attention Mechanisms is given in section \ref{sec:am}.
\vspace{-5pt}
\subsection{Deep Learning Based Image Super-Resolution} \label{sec:dl_sr}

Dong et al. \cite{dong2014learning} present one of the first work using CNN to tackle the SR task (i.e., SRCNN). The SRCNN receives an upsampled image as an input that cost extra computation. Later on, to address this drawback, FSRCNN \cite{dong2016accelerating} and ESPCN \cite{shi2016real} have been proposed to reduce the large computational and run time cost by upsampling the features near to the output of the network. This tactic leads results in efficient approaches with low memory compared to SRCNN. However, the entire performance could be reduced if there are not enough layers after the upsampling process. In addition, they cannot manage multi-scale training, as the size of the input image differs for each upsampling scale.

Even though the strength of deep learning shows up from deep layers, the above-mentioned methods are referred to as shallow network due to the training difficulties. Therefore, Kim et al. \cite{kim2016accurate} use residual learning to ease the training challenges and increase the depth of their network by adding 20 convolutional layers. Then, \cite{tai2017memnet} has proposed memory block in MemNet for deeper networks and solve the problem of long-term dependency with 84 layers. Thus, CNN-based SR approaches demonstrate that deeper networks with various types of skip connections show better performance. Thereby, Lim et al. \cite{lim2017enhanced} introduce EDSR by expanding the network size and enhancing the residual block by omitting the batch normalization from residual block. Zhang et al. \cite{zhang2018residual} propose RDN with residual and dense skip connections to fully use hierarchical features. Li et al. \cite{li2018multi} propose a network with more than 160 layers plus improved residual units. Despite of the fact that they achieve higher PSNR values, the number of parameters and operations are increased, which leads to high risk of overfitting and limits for real-world applications. 

% Following the work of \cite{kim2016deeply}, Tai et al. \cite{tai2017image} introduce DRRN to improve DRCN by fusing the recursive and residual blocks schemes.

% Later, DRCN \cite{kim2016deeply} introduces recursive learning to share parameters to handles deep network.

% So, building the lightweight network that can be trade off between network parameters and PSNR is still a challenging task.
% Wang et al. \cite{wang2018esrgan} proposed ESRGAN to use residual in residual dense block to improve the training stability and network size.

% by compressing pretrained networks or designing lightweight models. For example, Forrest et al. introduce squeezeNet \cite{iandola2016squeezenet} build upon the idea of AlexNet and achieves comparable results with $50\times$ fewer network parameters. Howard et al. \cite{howard2019searching} propose an efficient network by applying depth-wise separable and point-wise convolution to obtain comparable results.

\vspace{-5pt}
\subsection{Deep Learning Lightweights Super Resolution}\label{sec:dl_wsr}

In recent years the interest of building lightweight and efficient models has been increased in SISR to reduce the computational cost. Several lightweight networks have been introduced, such as SRCNN \cite{dong2014learning}, FSRCNN\cite{dong2016accelerating}, ESPCN\cite{shi2016real}, which were the first attempts, but they could not perform well. Later, Ahn et al. \cite{ahn2018fast} design a network that is suitable in the mobile scenario by implementing a cascade mechanism beyond a residual network (CARN), in order to obtain lightweight and improve reconstruction but it is at the cost of reduction of PSNR. Then, a neural architecture search (NAS)-based strategy has been also proposed in SISR to construct efficient networks---MoreMNA-S \cite{chu2019multi} and FALSR \cite{chu2019fast}. But due to limitation in strategy, the performance of these models are limited. Later, \cite{muqeet2020ultra} introduces MAFFSRN by proposing multi-attention blocks to improve the performance. Recently, LatticeNet \cite{liu2020residual} introduces an economical structure to adaptively combine Residual Blocks, which achieve good results. All these works suggest that the lightweight SR networks can keep a good trade-off between PSNR and parameters.            

% \cite{ahn2018fast} proposed an architecture that implements a cascading mechanism upon a residual network (CARN) that is applicable in the mobile scenario to achieve lightweight and efficient reconstruction but it is at the cost of reduction on PSNR. More recently, The NAS-based SR networks that tackles SISR with neural architecture search has also been proposed, such as MoreMNA-S \cite{chu2019multi} and FALSR \cite{chu2019fast}. However, due to the constraints of search space and strategy in NAS, the performance of NAS-based networks is also limited. All these works suggest that the lightweight SR networks can keep a good trade-off between reconstruction quality and the number of network parameters.}
% \vspace{-1pt}
\subsection{Attention Mechanism} \label{sec:am}

Attention can be described as a guide to bias the allocation of available computer resources to the most important informative elements of an input. Recently, some works have focused on attention mechanism for deep neural networks. Hu et al. \cite{hu2018squeeze} introduce squeeze-and-excitation (SE) block, a compact module to leverage the relationship between channels. Also, Woo et al. \cite{woo2018cbam} propose a Convolutional Block Attention Module (CBAM) to exploit the inner-spatial and inner-channel relationship of features to achieve a performance improvement in image classification. 

Recently, RCAN \cite{zhang2018image} designs a very deep network with a channel attention mechanism to enhance the reconstruction results by only considering inner-channel information, which call first-order statistics. In contrast, Dai et al. \cite{dai2019second} introduce the second-order attention network in order to explore more powerful feature expression. More recently, Li et al., \cite{liu2020residual} propose enhanced spatial attention (ESA) to make the residual features to be more focused on critical spatial contents. In the current work, motivated by attention mechanisms and considering that there are different types of information within and across feature space, which have different contributions for image SR, a Two-Fold Attention Mechanism is proposed that adaptively highlight the important information by considering both channel and spatial information to boost the performance of the network.

% ; this approach does not take into account spatial information, which is an important factor to recover the HR image. Furthermore, in these model \cite{hu2018squeeze, woo2018cbam, zhang2018image} channel information loss due to channel reduction. Having in mind these limitations, in the current approach a Two-Fold Attention Mechanism (TFAM) is proposed, which includes both channel and spatial information with keeping all the channel information to achieve better performance in the benchmark datasets and even in more realistic scenarios.

%-------------------------------------------------------------------------
\section{Multi-Path Residual Network }
% This section details the structure of proposed MPRNet model (see ).

\subsection{Network Structure} \label{sec:net_struc}
The proposed model (MPRNet -- Fig \ref{fig:arc}) consists of four different modules, namely, Shallow Feature Extraction (SFE); Residual Module that contains Residual Concatenation Blocks (RCBs); Feature Module that includes a Two-Fold Attention Mechanism (TFAM) and a Global Feature Extractor with a Long-Range Skip Connection; and the multi-scale UP-Net module at the end of network. Let’s consider $\{I_{LR}, I_{SR}\}$ as the input and output of the network respectively. The SFE is a Conv layer with a kernel size of $3\times3$, which can be formulated as follow:

\vspace{-10pt}
\begin{equation}
    \boldsymbol{H}_{SFE} = f_{SFE}(\boldsymbol{I}_{LR};W_c),
\end{equation}

\noindent where $f_{SFE}(\cdot)$ and $W_c$ indicates Conv operation and parameters applied on $I_{LR}$. $\boldsymbol{H}_{SFE}$ denotes the output of SFE, which later is used as the input to Residual Module. Lets $\boldsymbol{H}_{RM}^{i,j}$ be the output from the $i$-th Residual Concatenation Block (RCB) that has $j$-th inner Adaptive Residual Blocks (ARBs). The Residual Module can be defined as:

\vspace{-15pt}
\begin{equation}
\begin{aligned}
    \boldsymbol{H}_{RM} &= \\
    &f([\boldsymbol{H}_{SFE},..., \boldsymbol{H}_{RCB}^{i-1}(\boldsymbol{H}_{ARB}^{j-1,R};W_{c}^j), \boldsymbol{H}_{RCB}^{i}];W_c^i),
\end{aligned}
\end{equation}

\noindent where $\boldsymbol{H}_{RM}$ is the output of the Residual Module. Note that our RM contains multi-level learning connections followed by a $1\times1$ Conv layer to control the output after each block, which helps our model to quickly propagate information all over the network (lower to higher layers and vice-versa in term of back propagation) and also let the network to learn multi-level representations. So, $i$-th RCB can be defined as: 
\vspace{-7.0pt}
\begin{equation}
\begin{aligned}
    \boldsymbol{H}^{i}_{RCB} = f([\boldsymbol{H}_{ARB}^{j,R}, ..., \boldsymbol{H}^{j-1,R}_{ARB}(\boldsymbol{H}^{i-1};W_c^i)]; W_c^j).
\end{aligned}
\end{equation}

\begin{figure*}
        \centering
        \includegraphics[width=1.0\linewidth]{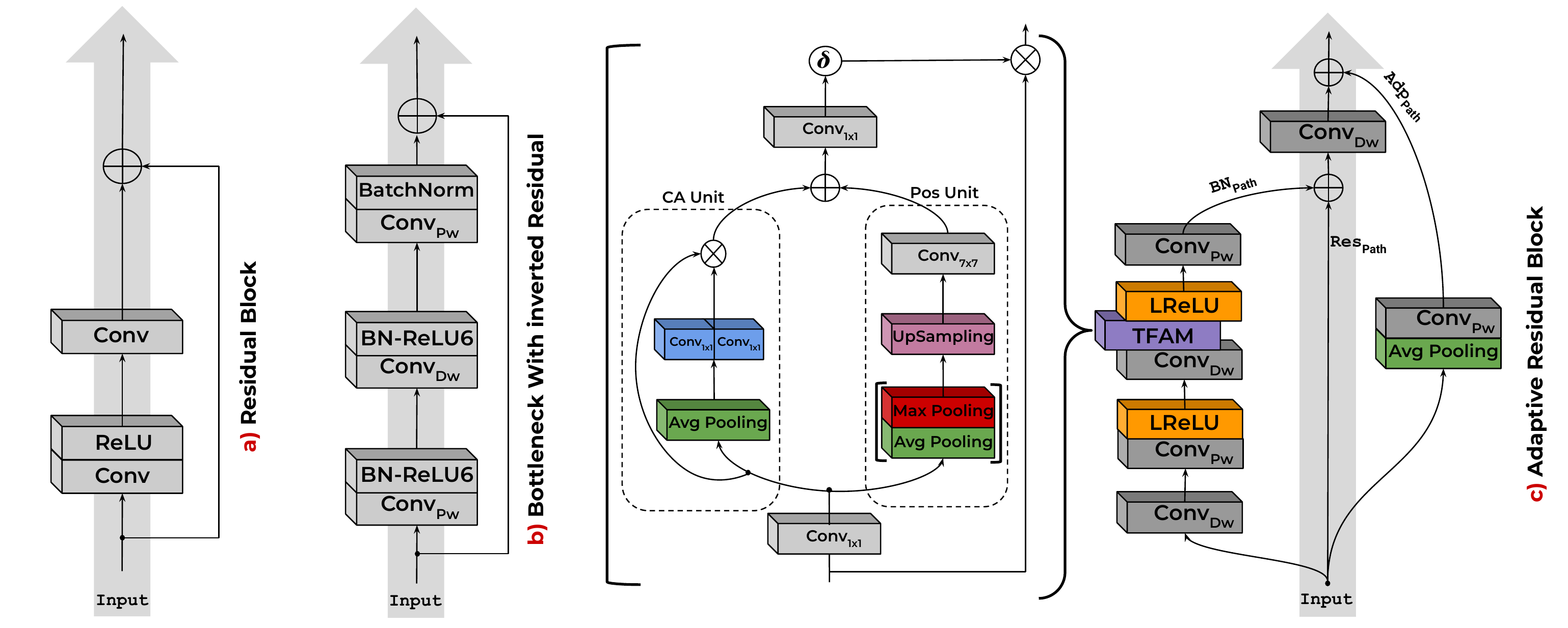}
        \caption{Illustrations of different structure of residual blocks: a) Residual block in EDSR \cite{lim2017enhanced}; b) Bottleneck with inverted residual from \cite{howard2019searching}; c) Proposed Adaptive Residual Block and Two-Fold Attettion Module.}
        \label{fig:arb}
\vspace{-18pt}
\end{figure*}

Then, the output of RM feed to the Feature Module by firstly refining the feature maps (i.e., re-calibrate) throughout the TFAM and then extracting more abstract features. Later, accumulate with LRSC to efficiently alleviate the gradient vanishing/exploding problems and make sure that network has access to unmodified information before UP-Net:

% \vspace{-20pt}
\begin{equation}
\begin{aligned}
   \boldsymbol{H}_{FM}= f_{GFE}(\boldsymbol{H}_{TFAM}(\boldsymbol{H}_{RM}; W_c); W_c) + \boldsymbol{H}_{LRC},
\end{aligned}
\end{equation}

\noindent where $\boldsymbol{H}_{TFAM}$ denotes our TFAM and $\boldsymbol{H}_{LRC}$ is Long-Range Residual Connection. The last stage is the Multi-Scale Up-Net Module to reconstruct the image from obtained feature-maps. The upsampling module is inspired by \cite{ahn2018fast} and followed by a Conv layer:
\vspace{-3.0pt}
\begin{equation}
   \boldsymbol{H}_{UP} = f^{\uparrow}_{pix}(\boldsymbol{H}_{FM}),
\end{equation}

\noindent where $f^{\uparrow}_{pix}(\cdot)$ indicates the Up-net module function and $\boldsymbol{H}_{FM}$ is the output of FM. The upsampled features are reconstructed with a Conv layer:
\begin{equation}
    \boldsymbol{I}_{SR} = f_{REC}(\boldsymbol{H}_{up}) = \boldsymbol{H}_{MPRNet}(\boldsymbol{I}_{LR}),
\end{equation}

\noindent where $f_{REC}(\cdot)$ and $\boldsymbol{H}_{MPRNet}(\cdot)$ denote the reconstruction layer and function of our MPRNet. In the next subsections, more details about the Adaptive Residual Block and Two Fold Attention Mechanism are given. 
% Finally, MPRNet optimized with $L1$ loss function. So, Given a training set $\{I^i_{lr}, I^i_{hr}\}^N_{i=1}$ that includes $N$ LR images and corresponding HR. Our objective is to minimize the $L1$ loss function
% \begin{equation}
% \begin{aligned}
%     L(\theta) = \frac{1}{N} \sum_{i=1}^{N} \ \lvert\lvert H_{MPRNet}(I_{LR}^i) - I^i_{HR} \rvert\rvert,
% \end{aligned}
% \end{equation}

% where $\theta$ indicates the parameter set. The loss function optimized by using ADAM optimizer (More details in section XX). 

% A novel structure has been proposed to be practical on low-capacity devices

\subsection{Adaptive Residual Block} \label{sec:LMB}
% In this section, details of the proposed Adaptive Residual Block (ARB) are given. 
This research focuses on designing a efficient and effective Residual Block based on Depthwise (Dw) and Pointwise (Pw) Convolutions for SISR. \cite{sandler2018mobilenetv2} introduced linear bottleneck with an inverted residual structure. However, this structure deliver chances of losing information and weaken the propagation capability of gradients across layers, due to gradient confusion arising from the narrowed feature space \cite{daquan2020rethinking, li2019hbonet}. Thus, we propose a novel Residual Block that mitigates the aforementioned issues; it is well-optimized especially for the SR tasks, called Adaptive Residual Block (ARB). Unlike \cite{sandler2018mobilenetv2}, ARB introduces new features and operations by proposing a multi learning pathways with a completely new structure. Each learning path is responsible to extract different kind of information before aggregation. So, the main part of network can have access to more rich information and performs notably well in noisy LR and generates more accurate SR image. The ARB consists of three different learning pathways that are detailed below. Fig \ref{fig:arb} shows each of the ARB components.

\textbf{Bottleneck Path}: We design our Bottleneck path (BN) based on the following insights: $i)$ Extract richer spatial information since spatial information is key importance in SR tasks; $ii)$ prevent very wide feature maps in the middle of the building block, which unavoidably growing the computational load of relevant layers; $iii)$ preserve the BN path low-cost and efficient. Thus, Dw Convolutions with small kernel size (3$\times$3) are chosen since they are lightweight and they can learn expressive features when conducted to the high dimensional space. So, we initiate the BN path by using a Dw convolution with kernel size 3$\times$3 towards the high dimensional features space to richer spatial information to be encoded and generate meaningful representations. Also, a Pw convolution is used after each Dw convolution in our design to produce new features by encoding the inter-channel information and reduce the computational cost. We shared the same number of channels and resolution along the BN path to prevent of sudden rise of computational burden in middle of the path. Furthermore, we conjunct our TFAM into the BN path after the second Dw convolution to spotlight the informative features along the channel and spatial axes. By doing so, the BN path is working with high dimensional features space, which makes the pathway efficient, low-cost, and well-focused on spatial context information compared to \cite{sandler2018mobilenetv2}.

\textbf{Adaptive Path}: It is proposed by taking the advantages of global average pooling accompanied by a 1$\times$1 Pw convolution. Average Pooling layers have been employed to take the average value of the features from the feature space to smooth and eliminate the noise from the LR image and reduce the dimensionality of each feature map but retains the important information to help the network to generate robust feature maps in challenging situations---noisy LR image. So, the network can generate a sharper and well-detailed SR image.
% \vspace{-3pt}
% \begin{equation}
%     \begin{aligned}
%         \boldsymbol{H}_{Adp} = f_p(\boldsymbol{P}_{avg}(\boldsymbol{H}^{i-1};W_c^{i,1});W_c^{i,2}),
%     \end{aligned}
% \end{equation}

% \noindent where $\boldsymbol{P}_{avg}(\cdot)$ denotes global average pooling.
% \vspace{50pt}
                               
% \vspace{100pt}
\textbf{Residual Path}: Unlike \cite{sandler2018mobilenetv2} that puts the residual path between narrowed feature space that cause gradient confusion, in our ARB, we place the residual path on the high dimensional representations to transfer more information from the bottom- to top-layers. Such structure facilitate the gradient propagation between multiple layers and help the network to optimize better during training. 

Thus, the information from BN- and Res-paths aggregate together, followed by another Dw convolution. We found out adding the Dw convolution before final aggregation with Adp path is essential for performance improvement since Dw encourage the network to learn more meaningful spatial information. Extensive experiments show that, our ARB is more beneficial than the existed ones for SISR tasks and improved the results with a large margin.

\vspace{-3.0pt}
\subsection{Two-Fold Attention Module} \label{sec:TFAM}

A novel Attention Mechanism (TFAM) has been proposed to boost the performance of our Adaptive Residual Block and refine the high-level information in the Feature Module (FM) by focusing on both channel and spatial information. The best way to amplify efficiency of ARB is through the union of the channel and spatial attention mechanism, since the residual features need to be well-focused on both information. In detail, TFAM is designed to focus on the important features on the channel information via CA unit and spotlight on the region of interest via Pos unit. Thus, each unit can learn ‘what’ and ‘where’ to attend in the channel and spatial axes respectively to recover edges and textures more accurately. As a result, TFAM works better than other attention mechanism \cite{woo2018cbam, hu2018squeeze, hu2019channel, liu2020residual} by emphasising informative features and reducing worthless ones. 

%It is worth to mentioned that several factors ought to be properly considered in the design of an attention module: $\textbf{i)}$ the attention module has to be low-cost because it will be injected into every residual block; $\textbf{ii)}$ less channel information have to be lose in CA unit to work properly; and $\textbf{iii)}$ a large spatial size is needed for Pos unit to work well in SR tasks. As illustrated in Fig \ref{fig:arb} TFAM initiates with 1$\times$1 Conv layer to reduce the channel dimensions, therefore the module can be greatly low-cost.

\textsc{\bf Channel Unit}. CA unit starts with an average pooling to exploit first-order statistics of features followed by two Conv layer, which they work side by side, each seeing half of the input channels, and producing half the output channels, and both subsequently concatenated to even have more low-cost unit. Thus, CA unit modulates features globally, where the summary statistics per channel are computed. Then, used to emphasize meaningful feature maps while redundant useless features are diminished. Especially, CA unit focuses on ‘what’ is meaningful given an input image.

% However, this kind of design for channel unit causes a restriction in losing channel information because of reduction. Particularly, first FC layer employed to reduce the input channel information by using reduction ratio ($C$ -$>$ $C/r$) and the second FC layer used to expands the reduced channel information to original size $C$. As a result, channel information will loss during these operations.  

% To overcome the mentioned issue, we propose a new channel unit for our TFAM by using only one FC layer with $C$ channel to preserves channel information instead of two FC layer and channel dimension reduction; result enhancing the performance of the model.  

% The average pooling is employed to produce channel-wise summary statistics $U^{ca}_{avg} \in \mathbb{R}^{C\times1\times1}$, on individual feature channels along spatial axis $H\times W$. In contrast to \cite{woo2018cbam}, max-pooling and sigmoid operations are omitted from CA unit and re-formulated to be suitable for low-capacity devices. Based on the experiments, max-pooling harms the overall performance of the model and sigmoid is costly to compute on edge devices: 
% \vspace{-15pt}
% \begin{equation}
% \begin{aligned}
%     \boldsymbol{U}_{ca}(\boldsymbol{H}^{i-1}, W^1) &= f(\boldsymbol{P}_{avg}(\boldsymbol{H}^{i-1};W^{i,1})), \\
% \end{aligned}
% \end{equation}
     
% \noindent where $H$ denotes the input to CA unit.$P_{avg}$ denotes average pooling.

\textsc{\bf Positional Unit}. Pos unit designed as a complementary unit to our CA unit. The feature map information is varied over spatial positions therefore, Pos unit concerns about the position of the informative part of the image and focuses on that region. Pos unit requires a large receptive field to work perfectly in SR tasks unlike the classification task. Thus, Average- and Max pooling operations with a large kernel size have been employed and then concatenated them to generate an efficient feature descriptor. afterward, an UpSampling layer is used to retrieve the spatial dimensions, which is followed by a Conv layer to generate a spatial attention map.

% In SR, it is necessary to pay attention to the regions that are difficult to recover, such as edge and texture regions. Thus, We proposed a pos unit, which works in parallel to our CA unit by producing pooled features map with both average and max-pooling operations ($U^{pos}_{avg,max} \in \mathbb{R}^{1\times H\times W}$) along the channel axes. Then, accumulate pooled feature by concatenation and convolved with a $5\times5$ Conv layer. Pooling operations spotlights informative fields along the channel axis to generate an effective feature descriptor. So, Pos unit can be formulated as:

% applying  is obtained by concatenating average and max-pooling and then convolved with a conv layer. Pooling operations spotlights informative fields along the channel axis to generate an effective feature descriptor. So, in order to calculate Pos unit, channel information of a feature map is accumulated; two $2D$ maps produced ($U^{pos}_{avg,max} \in \mathbb{R}^{1\times H\times W}$) by using of two pooling operations along the channel axis. So, Pos unit can be formulated as:

% \vspace{-10pt}
% \begin{equation}
% \begin{aligned}
%     &\boldsymbol{U}_{pos}(\boldsymbol{H}^{i-1}, W^1) = \\ & f([\boldsymbol{P}_{avg}(\boldsymbol{H}^{i-1};W^{i,1});\boldsymbol{P}_{max}(\boldsymbol{H}^{i-1};W^{i,1})];W_c),
% \end{aligned}
% \end{equation}

% \noindent where $f(\cdot)$ indicates a uses of Conv operations after computing and concatenate the pooled information. 
\begin{table*}[t!]
\caption{\small Comparison with lightweight SOTA methods on the Bicubic (\textbf{BI}) degradation for scale factors $[\times2, \times3, \times4]$. \textcolor{red}{\textbf{Red}} is the Best and \textcolor{blue}{\textbf{Blue}} is the second best performance. We assume that the generated SR image is $720P$ to calculate Multi-Adds (MAC).}
% \vspace{-5.0pt}
\begin{adjustbox}{width=1\textwidth}
\setlength\arrayrulewidth{1.0pt}
\begin{tabular}{|c|c|c|c|c|c|c|c|c|c|c|c||c|c|}
\hline
\begin{tabular}[c]{@{}c@{}}\multirow{2}{*}{} \\ \hline Params-MAC \\ \hline Dataset \end{tabular} &
\begin{tabular}[c]{@{}c@{}}\multirow{2}{*}{} \textbf{Methods} \\ \hline $\times$4 \\ \hline Scale \end{tabular} &
\begin{tabular}[c]{@{}c@{}}\multirow{2}{*}{} \textbf{VDSR} \cite{kim2016accurate} \\ \hline $655K-612.6G$ \\ \hline PSNR/SSIM \end{tabular} & \begin{tabular}[c]{@{}c@{}}\multirow{2}{*}{}\textbf{LapSRN}{\cite{lai2017deep}}\\ \hline  $813K-149.6G$\\ \hline PSNR/SSIM \end{tabular} &  \begin{tabular}[c]{@{}c@{}}\multirow{2}{*}{}\textbf{MemNet}{\cite{tai2017memnet}}\\ \hline  $677K-2662.4G$ \\ \hline PSNR/SSIM\end{tabular} &
\begin{tabular}[c]{@{}c@{}}\multirow{2}{*}{}\textbf{NLRN}{\cite{liu2018non}}\\ \hline $350K-32.5$ \\ \hline PSNR/SSIM\end{tabular} &
\begin{tabular}[c]{@{}c@{}}\multirow{2}{*}{}\textbf{SRFBN\_S}{\cite{li2019feedback}}\\ \hline $483K-119G$ \\ \hline PSNR/SSIM\end{tabular} &
\begin{tabular}[c]{@{}c@{}}\multirow{2}{*}{}\textbf{CARN}{\cite{ahn2018fast}}\\ \hline $1592K-90.9G$ \\ \hline PSNR/SSIM\end{tabular} & 
\begin{tabular}[c]{@{}c@{}}\multirow{2}{*}{}\textbf{CBPN} {\cite{zhu2019efficient}}\\ \hline $1197K-97.9G$  \\ \hline PSNR/SSIM\end{tabular} &
\begin{tabular}[c]{@{}c@{}}\multirow{2}{*}{}\textbf{OISR\_RK2\_s}{\cite{he2019ode}}\\ \hline $1540K-114.2G$ \\ \hline PSNR/SSIM\end{tabular} &
\begin{tabular}[c]{@{}c@{}}\multirow{2}{*}{}\textbf{MAFFSRN-L}{\cite{muqeet2020ultra}}\\ \hline $830K-38.6G$ \\ \hline PSNR/SSIM\end{tabular} &

\begin{tabular}[c]{@{}c@{}}\multirow{2}{*}{}\textbf{LatticeNet}{\cite{luolatticenet}}\\ \hline 777K-43.6G \\ \hline PSNR/SSIM\end{tabular} &
\begin{tabular}[c]{@{}c@{}}\multirow{2}{*}{}\textbf{MPRNet [Ours]}\\ \hline \textbf{538K-31.3G} \\ \hline PSNR/SSIM\end{tabular} \\ \hline \hline

Set5 & 
\begin{tabular}[c]{@{}c@{}}$\times2$\\ $\times3$\\ $\times4$\end{tabular} & 
\begin{tabular}[c]{@{}c@{}}$37.53/0.9587$\\ $33.66/0.9213$\\ $31.35/0.8838$\end{tabular} & 
\begin{tabular}[c]{@{}c@{}}$37.52/0.9590$\\ --------\\ $31.54/0.8850$\end{tabular} & 
\begin{tabular}[c]{@{}c@{}}$37.87/0.9597$\\ $34.09/0.9248$\\ $31.74/0.8893$\end{tabular} & 
\begin{tabular}[c]{@{}c@{}}$38.00/0.9603$\\ $34.27/0.9266$\\ $31.92/0.8916$\end{tabular} & 
\begin{tabular}[c]{@{}c@{}}$37.78/0.9597$\\ $34.20/0.9255$\\ $31.98/0.9594$\end{tabular} &
\begin{tabular}[c]{@{}c@{}}$37.76/0.9590$\\ $34.29/0.9255$\\ $32.13/0.8937$\end{tabular} & 
\begin{tabular}[c]{@{}c@{}}$37.90/0.9590$\\ --------\\ $32.21/0.8944$\end{tabular} & 
\begin{tabular}[c]{@{}c@{}}$37.90/0.9600$\\ $34.39/0.9273$\\ $32.21/0.8903$\end{tabular} & 
\begin{tabular}[c]{@{}c@{}}$38.07/0.9607$\\ $34.45/0.9277$\\ $32.20/0.8953$\end{tabular} & 
\begin{tabular}[c]{@{}c@{}}\textcolor{red}{$\boldsymbol{38.15/0.9610}$}\\ \textcolor{blue}{$34.53/0.9281$}\\ \textcolor{blue}{$32.30/0.8962$}\end{tabular} & 
\begin{tabular}[c]{@{}c@{}} \textcolor{blue}{$38.08/0.9608$} \\ \textcolor{red}{$\boldsymbol{34.57/0.9285}$}\\ \textcolor{red}{$\boldsymbol{32.38/0.8969}$}\end{tabular} \\ \hline

Set14 & 
\begin{tabular}[c]{@{}c@{}}$\times2$\\ $\times3$\\ $\times4$\end{tabular} & 
\begin{tabular}[c]{@{}c@{}}$33.03/09124$\\ $29.77/0.8314$\\ $28.01/0.7674$\end{tabular} & 
\begin{tabular}[c]{@{}c@{}}$33.08/0.9130$\\ --------\\ $28.19/0.7720$\end{tabular} & 
\begin{tabular}[c]{@{}c@{}}$33.28/0.9142$\\ $30.00/0.8350$\\ $28.26/0.7723$\end{tabular} &
\begin{tabular}[c]{@{}c@{}}$33.46/0.9159$\\ $30.16/0.8374$\\ $28.36/0.7745$\end{tabular} &
\begin{tabular}[c]{@{}c@{}}$33.35/0.9156$\\ $30.10/0.8350$\\ $28.45/0.7779$\end{tabular} &
\begin{tabular}[c]{@{}c@{}}$33.52/0.9166$\\ $30.29/0.8407$\\ $28.60/0.7806$\end{tabular} &
\begin{tabular}[c]{@{}c@{}}$33.60/0.9171$\\--------\\$28.63/0.7813$\end{tabular} & 
\begin{tabular}[c]{@{}c@{}}$33.58/0.9172$\\$30.33/0.8420$\\$28.63/0.7822$\end{tabular} & 
\begin{tabular}[c]{@{}c@{}}$33.59/0.9177$\\ $30.40/$\textcolor{blue}{$0.8432$}\\ $28.62/0.7822$\end{tabular} & 

\begin{tabular}[c]{@{}c@{}}\textcolor{blue}{$33.78/0.9193$}\\ \textcolor{blue}{$30.39$}$/0.8424$\\ \textcolor{blue}{$28.68/0.7830$}\end{tabular} & 

\begin{tabular}[c]{@{}c@{}} \textcolor{red}{$\boldsymbol{33.79/0.9196}$} \\ \textcolor{red}{$\boldsymbol{30.42/0.8441}$}\\ \textcolor{red}{$\boldsymbol{28.69/0.7841}$}\end{tabular}\\ \hline

B100 & 
\begin{tabular}[c]{@{}c@{}}$\times2$\\ $\times3$\\ $\times4$\end{tabular} &
\begin{tabular}[c]{@{}c@{}}$31.90/0.8960$\\ $28.82/0.7976$\\ $27.29/0.7251$\end{tabular} & 
\begin{tabular}[c]{@{}c@{}}$31.80/0.8950$\\ --------\\ $27.32/0.7280$\end{tabular} &
\begin{tabular}[c]{@{}c@{}}$32.08/0.8978$\\ $38.96/0.8001$\\ $27.40/0.7281$\end{tabular} &
\begin{tabular}[c]{@{}c@{}}$32.19/0.8992$\\ $29.06/0.8026$\\ $27.48/0.7306$\end{tabular} &
\begin{tabular}[c]{@{}c@{}}$32.00/0.8970$\\ $28.96/0.8010$\\ $27.44/0.7313$\end{tabular} &
\begin{tabular}[c]{@{}c@{}}$32.09/0.8978$\\ $29.06/0.8434$\\ $27.58/0.7349$\end{tabular} &
\begin{tabular}[c]{@{}c@{}}$32.17/0.8989$\\--------\\$27.58/0.7356$\end{tabular} & 
\begin{tabular}[c]{@{}c@{}}$32.18/0.8996$\\$29.10/0.8083$\\$27.58/0.7364$\end{tabular} &  

\begin{tabular}[c]{@{}c@{}}$32.23/$\textcolor{red}{$\boldsymbol{0.9005}$}\\ $29.13/$\textcolor{blue}{$0.8061$}\\ $27.59/$\textcolor{blue}{$0.7370$}\end{tabular} & 

\begin{tabular}[c]{@{}c@{}}$\textcolor{blue}{32.25}/\textcolor{red}{\boldsymbol{0.9005}}$\\ \textcolor{blue}{$29.15$}$/0.8059$\\ \textcolor{blue}{$27.62$}$/0.7367$\end{tabular} &

\begin{tabular}[c]{@{}c@{}} \textcolor{red}{$\boldsymbol{32.25}$}\textcolor{blue}{$/0.9004$}\\ \textcolor{red}{$\boldsymbol{29.17/0.8073}$}\\ \textcolor{red}{$\boldsymbol{27.63/0.7385}$}\end{tabular} \\ \hline

Urban100 &
\begin{tabular}[c]{@{}c@{}}$\times2$\\ $\times3$\\ $\times4$\end{tabular} &
\begin{tabular}[c]{@{}c@{}}$30.76/0.9140$\\ $27.14/0.8279$\\ $25.18/0.7524$\end{tabular} &
\begin{tabular}[c]{@{}c@{}}$30.41/0.9100$\\ --------\\ $25.21/0.7560$\end{tabular} & 
\begin{tabular}[c]{@{}c@{}}$31.31/0.9195$\\ $27.56/0.8376$\\ $25.50/0.7630$\end{tabular} &
\begin{tabular}[c]{@{}c@{}}$31.81/0.9249$\\ $27.93/0.8453$\\ $25.79/0.7729$\end{tabular} &
\begin{tabular}[c]{@{}c@{}}$31.41/0.9207$\\ $27.66/0.8415$\\ $25.71/0.7719$\end{tabular} &
\begin{tabular}[c]{@{}c@{}}$31.92/0.9256$\\ $28.06/0.8493$\\ $26.07/0.7837$\end{tabular} &
\begin{tabular}[c]{@{}c@{}}$32.14/0.9279$\\--------\\$26.14/0.7869$\end{tabular} &
\begin{tabular}[c]{@{}c@{}}$32.21/0.8950$\\$28.03/0.8544$\\$26.14/0.7874$\end{tabular} &
\begin{tabular}[c]{@{}c@{}}$32.38$\textcolor{blue}{$/0.9308$}\\ $28.26/$\textcolor{blue}{$0.8552$}\\ $26.16/$\textcolor{blue}{$0.7887$}\end{tabular} & 

\begin{tabular}[c]{@{}c@{}}\textcolor{blue}{$32.43$}$/0.9302$\\ \textcolor{blue}{$28.33$}$/0.8538$\\ \textcolor{blue}{$26.25$}$/0.7873$\end{tabular} & 
\begin{tabular}[c]{@{}c@{}} \textcolor{red}{$\boldsymbol{32.52/0.9317}$}\\ \textcolor{red}{$\boldsymbol{28.42/0.8578}$}\\ \textcolor{red}{$\boldsymbol{26.31/0.7921}$}\end{tabular} \\ \hline 
\end{tabular}
\end{adjustbox}
% \vspace{-0.65cm}
\label{tab:bicubic}
\end{table*}

\begin{table*}[t!]
\caption{\small Comparison with SOTA methods on challenging datasets ("\textbf{BD}" and "\textbf{DN}") for scale factor $\times3$. \textcolor{red}{\textbf{Red}} is the Best and \textcolor{blue}{\textbf{Blue}} is the second best performance.}
% \vspace{-5.0pt}
\begin{adjustbox}{width=1\textwidth}
\setlength\arrayrulewidth{1.0pt}
\begin{tabular}{|c|c|c|c|c|c|c|c|c|c|c||c|}
\hline
Dataset & \begin{tabular}[c]{@{}c@{}} \textbf{Methods}\\ \hline Degradation\end{tabular} & \begin{tabular}[c]{@{}c@{}} 
\textbf{Bicubic}\\ \hline PSNR/SSIM\end{tabular} & \begin{tabular}[c]{@{}c@{}} \textbf{SPMSR}\cite{peleg2014statistical}\\ \hline PSNR/SSIM\end{tabular} & \begin{tabular}[c]{@{}c@{}} \textbf{SRCNN}\cite{dong2014learning}\\ \hline PSNR/SSIM\end{tabular} & \begin{tabular}[c]{@{}c@{}} \textbf{FSRCNN}\cite{dong2016accelerating}\\ \hline PSNR/SSIM\end{tabular} & \begin{tabular}[c]{@{}c@{}} \textbf{VDSR}\cite{kim2016accurate}\\ \hline PSNR/SSIM\end{tabular} & \begin{tabular}[c]{@{}c@{}} \textbf{IRCNN\_G}\cite{zhang2017learning}\\ \hline PSNR/SSIM\end{tabular} & \begin{tabular}[c]{@{}c@{}} \textbf{IRCNN\_C}\cite{zhang2017learning}\\ \hline PSNR/SSIM\end{tabular} & \begin{tabular}[c]{@{}c@{}} \textbf{SRMD(NF)}\cite{tong2017image}\\ \hline PSNR/SSIM\end{tabular} & \begin{tabular}[c]{@{}c@{}} \textbf{RDN}\cite{zhang2018residual}\\ \hline PSNR/SSIM\end{tabular} & \begin{tabular}[c]{@{}c@{}} \textbf{MPRNet [Ours]}\\ \hline PSNR/SSIM\end{tabular} \\ \hline \hline
Set5 & \begin{tabular}[c]{@{}c@{}} \textbf{BD}\\ \textbf{DN}\end{tabular} & \begin{tabular}[c]{@{}c@{}}$28.34/0.8161$\\ $24.14/0.5445$\end{tabular} & \begin{tabular}[c]{@{}c@{}}$32.21/0.9001$\\ ----\end{tabular} & \begin{tabular}[c]{@{}c@{}}$31.75/0.8899$\\ $27.04/0.7638$\end{tabular} & \begin{tabular}[c]{@{}c@{}}$26.58/0.8224$\\ $24.28/0.7124$\end{tabular} & \begin{tabular}[c]{@{}c@{}}$33.29/0.9139$\\ $27.42/0.7372$\end{tabular} & \begin{tabular}[c]{@{}c@{}}$33.38/0.9182$\\ $24.85/0.7205$\end{tabular} & \begin{tabular}[c]{@{}c@{}}$29.55/0.8246$\\ $26.18/0.7430$\end{tabular} & \begin{tabular}[c]{@{}c@{}}$34.09/0.9242$\\ $27.74/0.8026$\end{tabular} & 
\begin{tabular}[c]{@{}c@{}}\textcolor{red}{$\boldsymbol{34.57/0.9280}$}\\ \textcolor{blue}{$28.46/0.8151$}\end{tabular} & 

\begin{tabular}[c]{@{}c@{}}\textcolor{red}{$\boldsymbol{34.57}$}\textcolor{blue}{$/0.9278$}\\ \textcolor{red}{$\boldsymbol{28.54/0.8175}$}\end{tabular}\\ \hline

Set14 & \begin{tabular}[c]{@{}c@{}}\textbf{BD}\\ \textbf{DN}\end{tabular} & \begin{tabular}[c]{@{}c@{}}$26.12/0.7106$\\ $23.14/0.4828$\end{tabular} & \begin{tabular}[c]{@{}c@{}}$28.89/0.8105$\\ ----\end{tabular} & \begin{tabular}[c]{@{}c@{}}$28.64/0.7997$\\ $25.56/0.6592$\end{tabular} & \begin{tabular}[c]{@{}c@{}}$24.86/0.7246$\\ $23.25/0.5956$\end{tabular} & \begin{tabular}[c]{@{}c@{}}$29.58/0.8259$\\ $25.60/0.6706$\end{tabular} & \begin{tabular}[c]{@{}c@{}}$29.73/0.8292$\\ $23.84/0.6091$\end{tabular} & \begin{tabular}[c]{@{}c@{}}$27.33/0.7135$\\ $24.68/0.6300$\end{tabular} & \begin{tabular}[c]{@{}c@{}}$30.11/0.8364$\\ $26.13/0.6974$\end{tabular} & \begin{tabular}[c]{@{}c@{}}\textcolor{red}{$\boldsymbol{30.53/0.8447}$}\\ \textcolor{red}{$\boldsymbol{26.60/0.7101}$}\end{tabular} & \begin{tabular}[c]{@{}c@{}}\textcolor{blue}{$30.47/0.8427$}\\ \textcolor{blue}{$26.25/0.6954$}\end{tabular} \\ \hline

B100 & \begin{tabular}[c]{@{}c@{}}\textbf{BD}\\ \textbf{DN}\end{tabular} & \begin{tabular}[c]{@{}c@{}}$26.02/0.6733$\\ $22.94/0.4461$\end{tabular} & \begin{tabular}[c]{@{}c@{}}$28.13/0.7740$\\ ----\end{tabular} & \begin{tabular}[c]{@{}c@{}}$27.33/0.7500$\\ $25.45/0.6198$\end{tabular} & \begin{tabular}[c]{@{}c@{}}$24.15/0.6728$\\ $23.95/0.5695$\end{tabular} & \begin{tabular}[c]{@{}c@{}}$28.61/0.7900$\\ $25.22/0.6271$\end{tabular} & \begin{tabular}[c]{@{}c@{}}$28.65/0.7922$\\ $23.89/0.5688$\end{tabular} & \begin{tabular}[c]{@{}c@{}}$26.46/0.6572$\\ $24.52/0.5850$\end{tabular} & \begin{tabular}[c]{@{}c@{}}$28.98/0.8009$\\ $25.64/0.6495$\end{tabular} & \begin{tabular}[c]{@{}c@{}}\textcolor{red}{$\boldsymbol{29.23/0.8079}$}\\ \textcolor{blue}{$25.93/0.6573$}\end{tabular} & \begin{tabular}[c]{@{}c@{}}\textcolor{blue}{$29.19/0.8062$}\\ \textcolor{red}{$\boldsymbol{25.95/0.6616}$}\end{tabular} \\ \hline

Urban100 & \begin{tabular}[c]{@{}c@{}}\textbf{BD}\\ \textbf{DN}\end{tabular} & \begin{tabular}[c]{@{}c@{}}$23.20/0.6661$\\ $21.63/0.4701$\end{tabular} & \begin{tabular}[c]{@{}c@{}}$25.84/0.7856$\\ ----\end{tabular} & \begin{tabular}[c]{@{}c@{}}$25.19/0.7591$\\ $23.59/0.6580$\end{tabular} & \begin{tabular}[c]{@{}c@{}}$22.95/0.6836$\\ $21.74/0.5724$\end{tabular} & \begin{tabular}[c]{@{}c@{}}$26.68/0.8019$\\ $23.33/0.6579$\end{tabular} & \begin{tabular}[c]{@{}c@{}}$26.77/0.8154$\\ $21.96/0.6018$\end{tabular} & \begin{tabular}[c]{@{}c@{}}$24.89/0.7172$\\ $22.63/0.6205$\end{tabular} & \begin{tabular}[c]{@{}c@{}}$27.50/0.8370$\\ $24.28/0.7092$\end{tabular} & \begin{tabular}[c]{@{}c@{}}\textcolor{red}{$\boldsymbol{28.46/0.8581}$}\\ \textcolor{blue}{$24.92/0.7362$}\end{tabular} &  
\begin{tabular}[c]{@{}c@{}}\textcolor{blue}{$28.31/0.8538$}\\ \textcolor{red}{$\boldsymbol{25.00/0.7406}$}\end{tabular}\\ \hline
\end{tabular}
\end{adjustbox}
% \vspace{-10pt}
\label{tab:BN}
\end{table*}
Finally, highlighted information from both units aggregated together followed a 1$\times$1 Conv layer and a sigmoid operation to firstly, recover the channel dimensions and then generate the final attention mask. Also, a residual connection used to transfer HR features to the end of module.

% \begin{equation}
% \begin{aligned}
%     \boldsymbol{H}_{TFAM} = \delta (\boldsymbol{U}_{ca} + \boldsymbol{U}_{pos}) \times H^{i-1},
% \end{aligned}
% \end{equation}

% \noindent where the $\delta(\cdot)$ is a sigmoid function. The details of our TFAM are illustrated in Fig.\ref{fig:arb}. 

% In section \ref{sec:study} an exhaustive ablation study on the impact of both, ARB and TFAM, is given in detail.
% effect of each component of ARB and impact of different pooling layer for both units of the TFAM on the SR results is analyzed in detail. 
% In the experimental result section an exhaustive ablation study is presented, where the effect of each component on the reconstruction results is analyzed in detail. 

% \input{files/tables/Tabel_BI_M1.tex}
\vspace{-7pt}
\section{Experimental Results}
% This section presents an extensive evaluation of the proposed MPRNet on series of benchmark datasets. Also, comparisons with a large set of SOTA algorithms are provided.

% n this section, we evaluate the performance of our modelon a series of benchmark datasets.  In addition, we providecomparison with state-of-the-art algorithms

\vspace{-3pt}
\subsection{Setting}
\textsc{\bf Datasets \& Evaluation Protocol.} Following previous works \cite{liu2020residual, dai2019second}, we use $DIV2K$ \cite{timofte2017ntire} dataset to train ($800$ images) and validate ($100$ images) our model. The proposed model is evaluated with the standard benchmark datasets, namely, $Set5$ \cite{bevilacqua2012low}, $Set14$ \cite{zeyde2010single}, $B100$ \cite{martin2001database}, and $Urban100$ \cite{huang2015single}. Two widely used quantitative metrics have been considered to measure its performance: PSNR and SSIM \cite{wang2004image}, computed between the obtained images and the corresponding ground truths. Both metrics are computed on the $Y$ channel in the $YCbCr$ space. 

\textsc{\bf Degradation Models.} Following the work of \cite{zhang2018residual}, three different degradation models created to simulate LR images and make fair comparisons with available methods. Firstly, a bicubic (BI) down-sampling dataset with scaling factors [$\times2$, $\times3$, $\times4$] has been created. Blur-Down-sampled (BD) is the second one to blur and down-sample HR images with a Gaussian kernel 7$\times$7, and $\sigma=1.6$. Then, images are down-sampled with scaling factor $\times3$. Aside from the BD, a more challenging model has been created, referred to as (DN). DN degradation model is down-sampling HR images with bicubic followed by adding $30\%$ Gaussian noise.

\textsc{\bf Training Details.} In the training stage, RGB input patches are used with size of 64$\times$64 from each of the randomly selected $64$ LR training images. Patches are augmented by random horizontally flips and 90 degree rotation. AdamP \cite{heo2020slowing} optimizer has been employed. The initial learning rate set to $10^{-3}$ and its halved every $4\times10^{5}$ steps. $L1$ is used as loss function to optimize the model. The PyTorch framework is used.

\vspace{-0.3 cm}
\setlength\tabcolsep{0.7pt}
\begin{figure}[b!]
\centering
\tiny
\begin{tabular}{ccccc}
\multirow{-6.715}{*}{\adjustbox{valign=b}{\includegraphics[width=.3\linewidth, height=2.99cm]{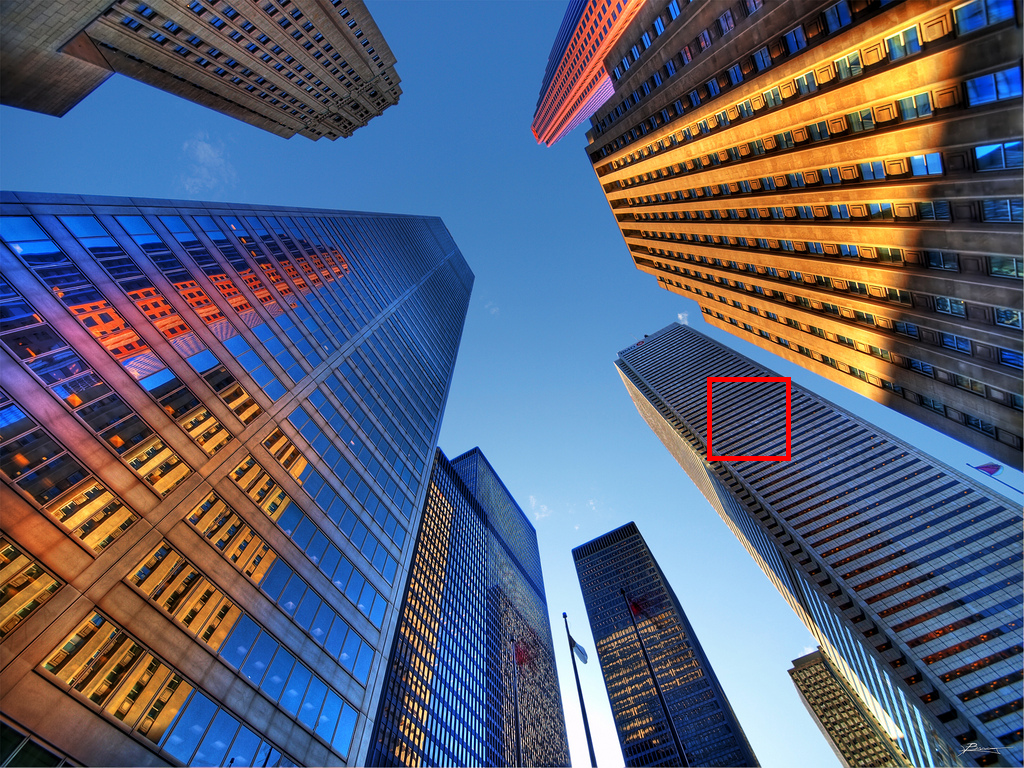}}} 
 & \includegraphics[width=.15\linewidth, height=1.354cm]{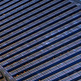} & \includegraphics[width=.15\linewidth, height=1.354cm]{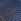} & \includegraphics[width=.15\linewidth, height=1.354cm]{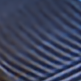} & \includegraphics[width=.15\linewidth, height=1.354cm]{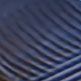} \\
 & HR & Bicubic & VDSR & MemNet \\
 & \includegraphics[width=.15\linewidth, height=1.354cm]{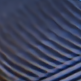} & \includegraphics[width=.15\linewidth, height=1.354cm]{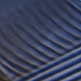} & \includegraphics[width=.15\linewidth, height=1.354cm]{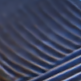} & \includegraphics[width=.15\linewidth, height=1.354cm]{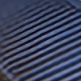} \\
 img012 from Urban100 & LapSRN & CARN & SRFBN-S & MPRNet(Ours) \\
\multirow{-6.715}{*}{\adjustbox{valign=b}{\includegraphics[width=.3\linewidth, height=2.99cm]{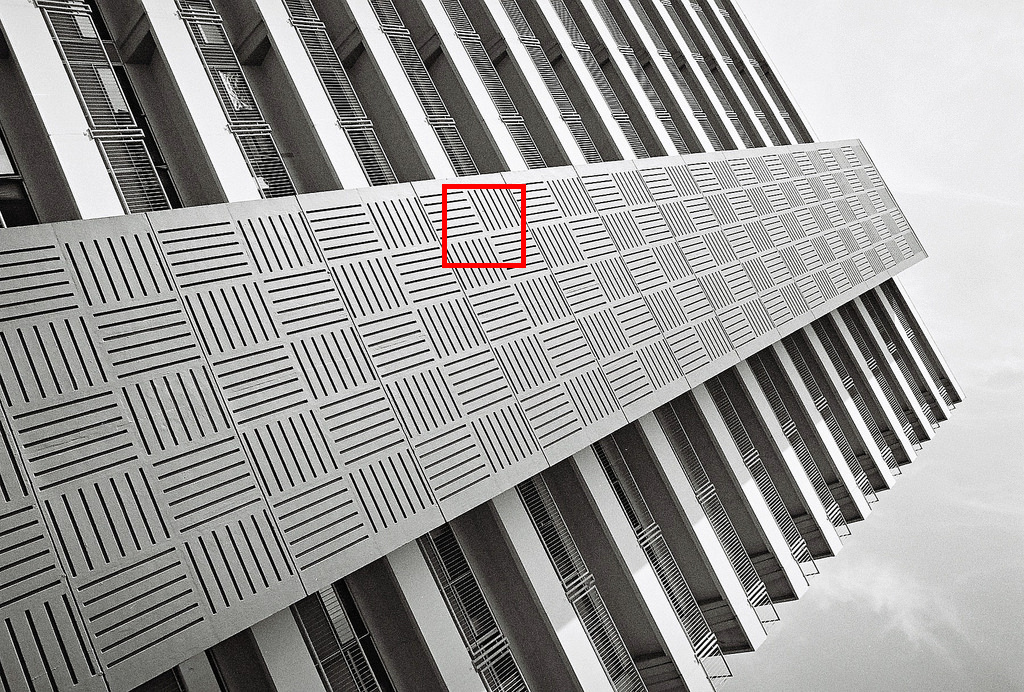}}} 
 & \includegraphics[width=.15\linewidth, height=1.354cm]{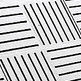} & \includegraphics[width=.15\linewidth, height=1.354cm]{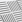} & \includegraphics[width=.15\linewidth, height=1.354cm]{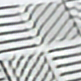} & \includegraphics[width=.15\linewidth, height=1.354cm]{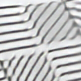} \\
 & HR & Bicubic & VDSR & MemNet \\
 & \includegraphics[width=.15\linewidth, height=1.354cm]{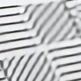} & \includegraphics[width=.15\linewidth, height=1.354cm]{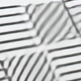} & \includegraphics[width=.15\linewidth, height=1.354cm]{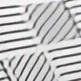} & \includegraphics[width=.15\linewidth, height=1.354cm]{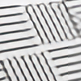} \\
img092 from Urban100 & LapSRN & CARN & SRFBN-S & MPRNet(Ours) \\
\end{tabular}
\caption{Qualitative results on \textbf{BI} degradation dataset with scale factor $\times$4.}
\label{tab:Bi_fig}
\end{figure}
\setlength\tabcolsep{0.4pt}
\begin{figure}[b!]
\centering
\tiny
\begin{tabular}{ccccc}
\multirow{-6.715}{*}{\adjustbox{valign=b}{\includegraphics[width=.3\linewidth, height=2.99cm]{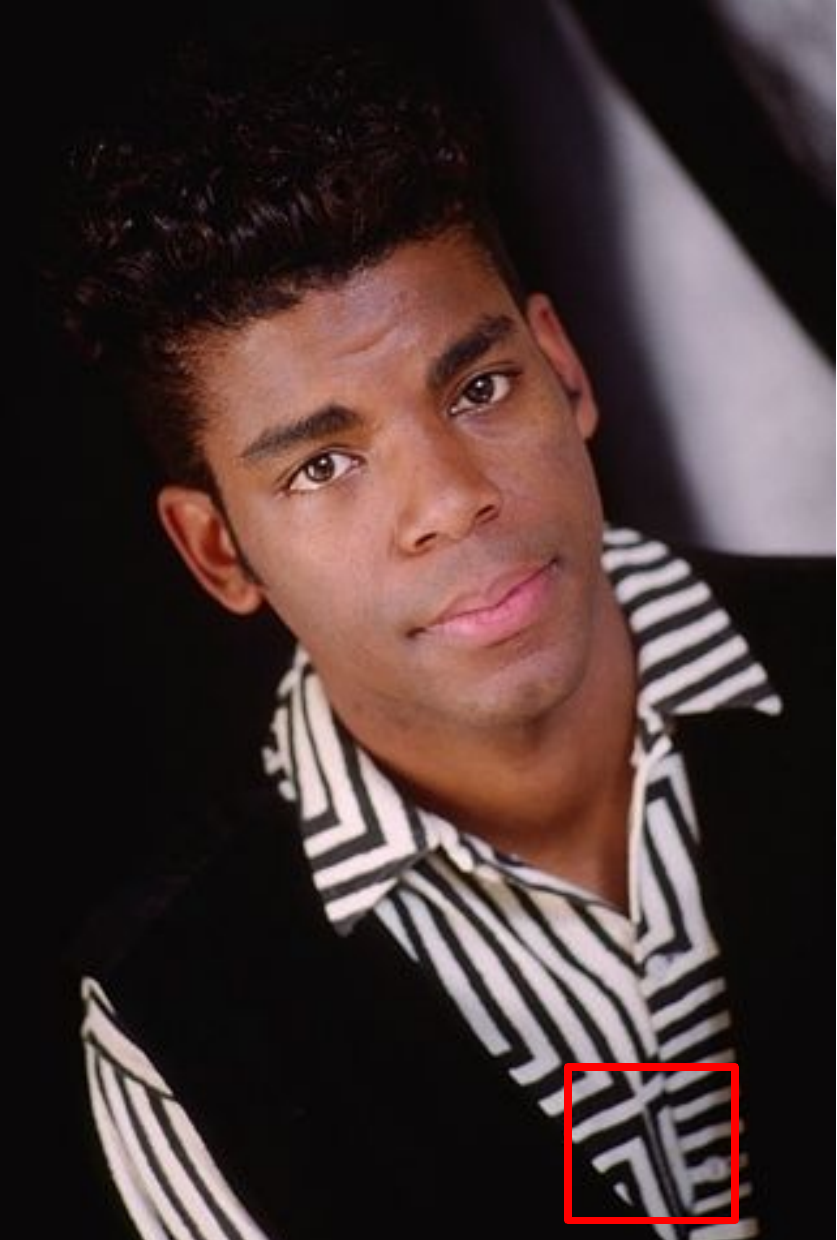}}} 
 & \includegraphics[width=.15\linewidth, height=1.354cm]{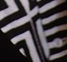} & \includegraphics[width=.15\linewidth, height=1.354cm]{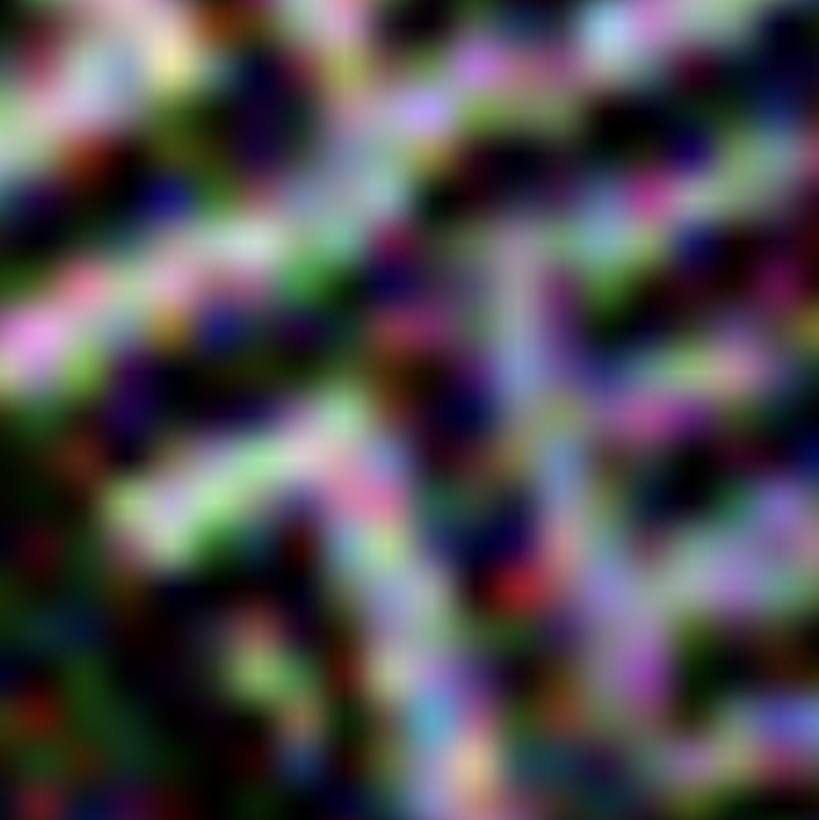} & \includegraphics[width=.15\linewidth, height=1.354cm]{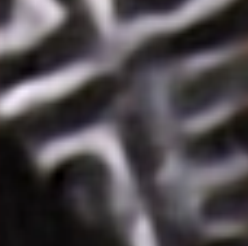} & \includegraphics[width=.15\linewidth, height=1.354cm]{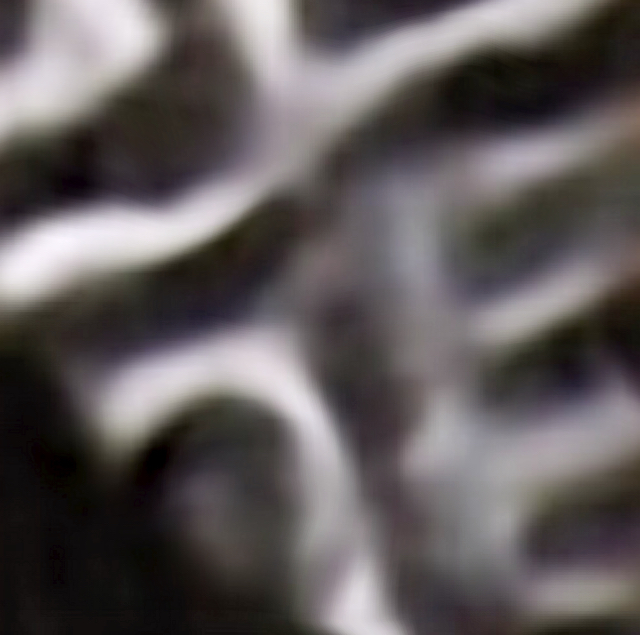} \\
 & HR & Bicubic & SRCNN & FSRCNN \\
 & \includegraphics[width=.15\linewidth, height=1.354cm]{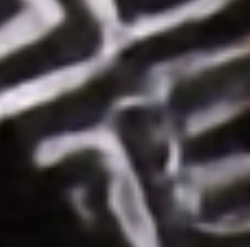} & \includegraphics[width=.15\linewidth, height=1.354cm]{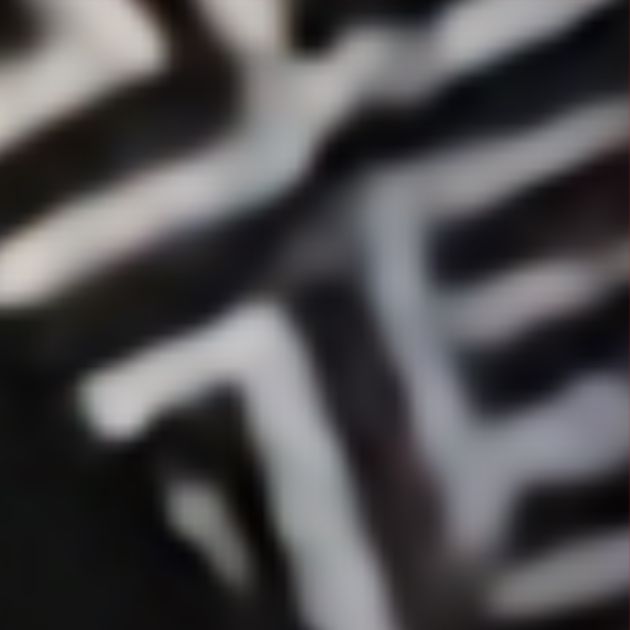} & \includegraphics[width=.15\linewidth, height=1.354cm]{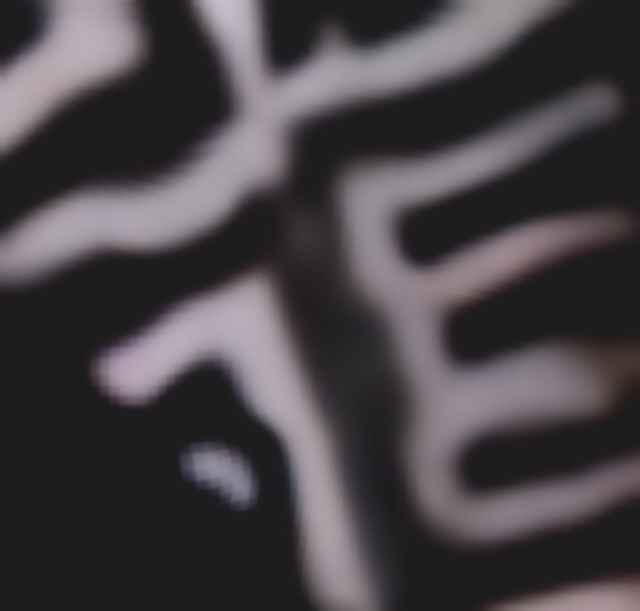} & \includegraphics[width=.15\linewidth, height=1.354cm]{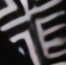} \\
 img063 from B100 & VDSR & IRCNN\_C & SRMD & MPRNet(Ours) \\
\multirow{-6.715}{*}{\adjustbox{valign=b}{\includegraphics[width=.3\linewidth, height=2.99cm]{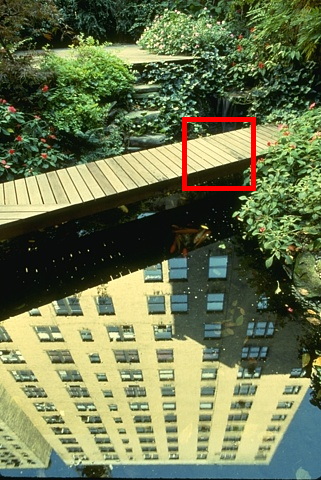}}} 
 & \includegraphics[width=.15\linewidth, height=1.354cm]{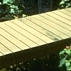} & \includegraphics[width=.15\linewidth, height=1.354cm]{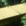} & \includegraphics[width=.15\linewidth, height=1.354cm]{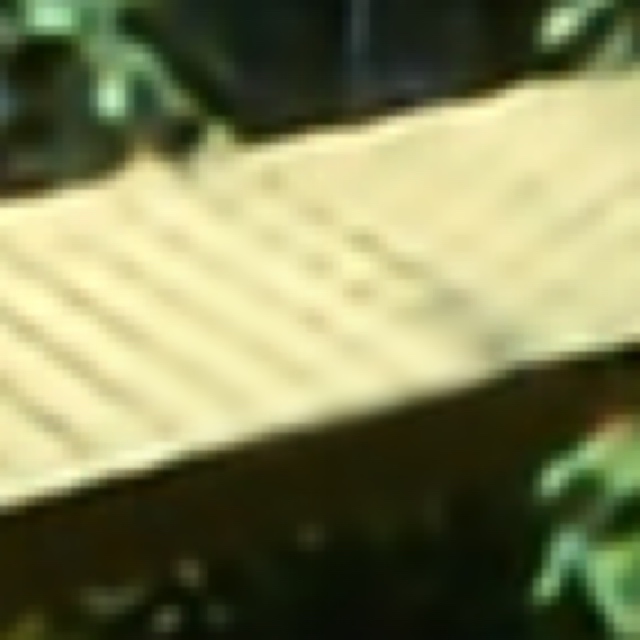} & \includegraphics[width=.15\linewidth, height=1.354cm]{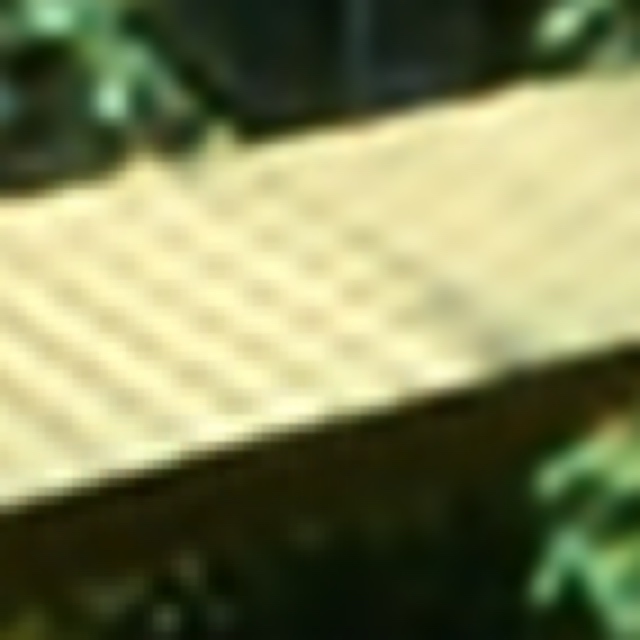} \\
 & HR & Bicubic & SRCNN & FSRCNN \\
 & \includegraphics[width=.15\linewidth, height=1.354cm]{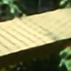} & \includegraphics[width=.15\linewidth, height=1.354cm]{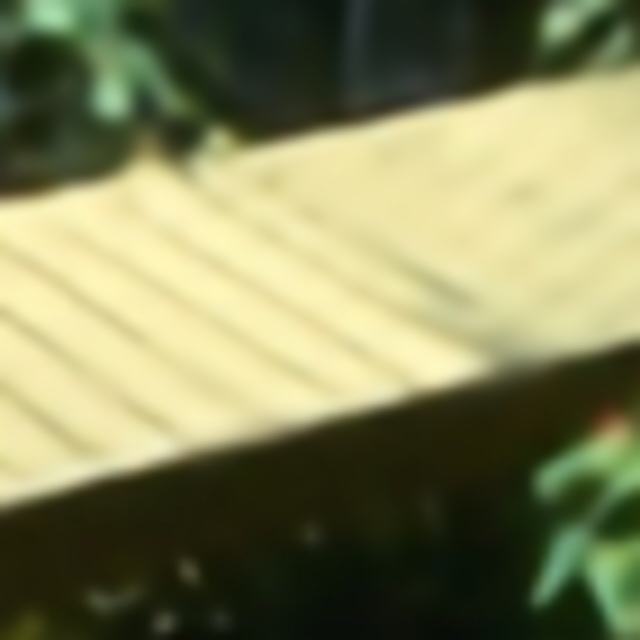} & \includegraphics[width=.15\linewidth, height=1.354cm]{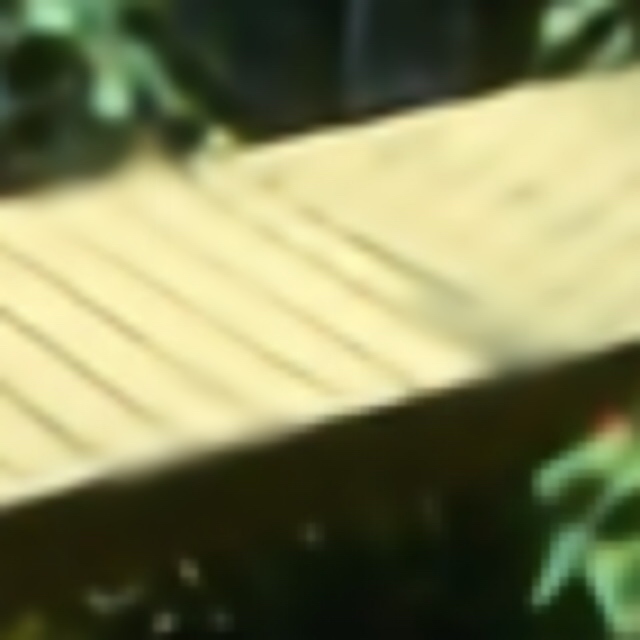} & \includegraphics[width=.15\linewidth, height=1.354cm]{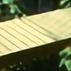} \\
img021 from B100 & VDSR & IRCNN\_G & SRMDNF & MPRNet(Ours) \\
\end{tabular}
\caption{Qualitative results on \textbf{DN} and \textbf{BD} degradation datasets with a scale factor $\times$3.}

\label{tab:BD_fig}
\end{figure}

%  First row corresponds to a SR result from \textbf{DN} degradation dataset. The second row corresponds to SR result from \textbf{BD} degradation dataset.

% MAC is computed by presuming the size of HR image as $1280\times720$.
\vspace{1pt}

\subsection{Comparison with state-of-the-art Methods}
% \vspace{-0.28cm}
% This section presents comparisons with state-of-the-art (SOTA) SR methods. Furthermore, the number of network parameters and the number of multiply-accumulate operations (MAC) are depicted in Table \ref{tab:bicubic} and Table \ref{tab:bicubic_M1} for more fair comparisons. The proposed model is compared with the recent lightweight SOTA methods and also due to its remarkable achievement, it is possible to compare it with models that are even ten times heavier; even in those scenarios the proposed MPRNet approach can achieves better or competitive results among all of them. This shows that the MPRNet model has a good balance between the number of parameters and the efficiency of the restoration.

% \input{files/tables/params.tex}

% \textbf{\bf Width Multiplier}. Similar to \cite{howard2019searching}, a width multiplier $\alpha$ has been employed to even make the model more lightweight with small reduction in reconstruction results. $\alpha$ control the input width of a layer, which can be $[1, 0.75, 0.5, 0.25]$. $\alpha=1$ is the baseline model and by decreasing $\alpha$ the computational cost and network parameters can be quadratically reduced by almost $\alpha^2$. It is worth to mention that the proposed model, with $295.7K$ ($\alpha=0.25$), can roughly outperform all the lightweight SOTA methods with $5\times$ heavier computation. In Table \ref{tab:NumParams}, the effect of width multiplier on network parameters and PSNR are shown.
% \vspace{-10pt}

\subsubsection{Results with BI Degradation}
% Table \ref{tab:bicubic} shows the quantitative comparisons on benchmark datasets for scale factor [$\times2$, $\times3$, and $\times4$]

%  and when it compared with very deep and computational expensive networks such as MSRN ($6000K$ parameters and $160$), MPRNet with only $526K$ parameters can still achieve almost better results on all benchmark datasets for all scale factors (\ref{fig:ParamsMac})

% Bear in mind that, in this table, we only consider the models that have roughly similar number of parameters as ours.
Table \ref{tab:bicubic} presents comparisons between the proposed MPRNet and $10$ most recent lightweight SOTA models on BI degradation model for scale factor [$\times2$, $\times3$, and $\times4$] to verify the effectiveness of our MPRNet (we exclude some lightweight methods \cite{dong2014learning, dong2016accelerating, tai2017image, kim2016deeply, wang2016studying, shi2016real} from table \ref{tab:bicubic} since their results are worse than MemNet). Table \ref{tab:bicubic} also contains the number of parameters and operations to show the model complexity. In almost all the cases, our MPRNet achieves superior results among all the aforementioned approaches. MPRNet performs especially well on Urban100. This is particularly because the Urban100 includes rich structured contents and our model can consistently accumulate these hierarchical features to form of more representative features and well-focused on spatial context information. This characteristic can be confirmed by our MPRNet SSIM scores, which focuses on the visible structures in the image. In Fig. \ref{tab:Bi_fig} a couple of qualitative results on scale factor $\times4$ are depicted. The proposed MPRNet can generally yield to more precise details. In both images in Fig. \ref{tab:Bi_fig}, the texture direction of the reconstructed images from all compared methods is completely wrong. However, results from the proposed MPRNet makes full use of the abstract features and recover images accurately similar to ground truth texture.
% \footnote{Results in Table 1 and 2 are from the most recent published papers \cite{he2019ode,ahn2018fast}.}

% The proposed MPRNet model outperforms all the lightweight SOTA approaches with large margin on all benchmark datasets. In Table \ref{tab:bicubic_M1}, comparisons with the heavier computational SOTA methods (Up to $Parameters < 6000K$) are given. MPRNet can also achieve better results than the very deep and computational expensive methods. For instance, the proposed MPRNet, with just $508K$ params, can achieves better SR results in almost all the benchmark datasets when compared to MSRN\cite{li2018multi} with $6078K$ parameters and $160$ layers ($\times12$ heavier computation). 

% (to be noted, RDN achieves sightly better results in BI degradation but not in the challenging datasets as shown in table \ref{tab:BN})}

%  and 

% effectiveness of our model in challenging datasets and when compared to computational expensive model

% The performance of MPRNet ,together with SOTA methods, are presented in Table \ref{tab:BN}, when BD and DN benchmark datasets are considered.

\vspace{-10pt}
\subsubsection{Results with BD and DN Degradation Models}

In Table \ref{tab:BN}, the performance of MPRNet on BD and DN benchmark datasets, together with SOTA methods, are presented. Due to degradation mismatch, SRCNN, FSRCNN, and VDSR for both BD and DN have been re-trained. As can be appreciated, MPRNet achieves remarkable results over all the lightweight SOTA models on challenging benchmark datasets. RDN \cite{zhang2018residual} also listed as a high-capability model to show the superior performance of MPRNet compared to very costly model in the BD and DN datasets. RDN performs sightly better in some BD datasets but not in DN datasets. Obviously, this result was expected since RDN is very expensive compared to low-cost MPRNet (it is almost $\times44$ more costly). Fig \ref{tab:BD_fig} depicts some visual results on both challenging BD and DN benchmark datasets. As can be appreciated the MPRNet with the help of the proposed TFAM performs better in comparison with SOTA methods in terms of producing more convincing results by cleaning off noise and blurred regions from SR images, which results in a sharper SR image with fine details.\footnote{Additional analyses (such as Inference time, Memory consumption, and etc.) and more visual results can be found in supplementary material.} 
% \vspace{-10pt}

% \subsection{Memory Complexity Analysis}
% Figure \ref{fig:ParamsMac} presents comparative results in terms of number of parameters vs PSNR and MAC vs PSNR. The proposed model can achieve the best results against the all lightweight networks and it can achieves competitive results to compare with the heavy computational methods. For example, our proposed MPRNet ($461K-106G$) can achieve better results with only needed of 8\% of MSRN ($5930K-1365.4G$) parameters and operations. Therefore, the proposed model can prove that it is well-balanced in terms of number of parameters, MAC, and reconstruction results (PSNR); and more efficient compare to other SOTA.  

%\vspace{-5pt}

\begin{table*}[]
\caption{Effect of Attention Mechanisms and proposed Adaptive Residual Block on SOTA models. The best \textbf{PSNR} (dB) are highlighted.}
% \vspace{-20pt}
\label{tab:channel_com}
\begin{center}
\resizebox{\textwidth}{!}{
\begin{tabular}{|l|c|c|ccc|c|c|c|ccc|c|c|c|ccc|c|}
\hline
% \multicolumn{19}{c}{Comparison of SISR state-of-the-art approaches with respect to different Attention Mechanisms and Adaptive Residual Block } \\ 
% \hline
\multicolumn{1}{|c|}{\multirow{2}{*}{\textbf{Name}}}                             &  & \multicolumn{5}{c|}{\textbf{EDSR}}                                                                                                                                                                                                                                                    &           & \multicolumn{5}{c|}{\textbf{RCAN}}                                                                                                                                                                                                                                                    &           & \multicolumn{5}{c|}{\textbf{MSRN}}                                                                                                                                                                                                                                                    \\ \cline{3-7} \cline{9-13} \cline{15-19} 
\multicolumn{1}{|c|}{}                                                  &  & \textbf{Baseline}                                     & \multicolumn{3}{c|}{\textbf{Attention Modules}}                                                                                                                       & \textbf{ResBlock}                                     & \textbf{} & \textbf{Baseline}                                     & \multicolumn{3}{c|}{\textbf{Attention Modules}}                                                                                                                       & \textbf{ResBlock}                                     & \textbf{} & \textbf{Baseline}                                     & \multicolumn{3}{c|}{\textbf{Attention Modules}}                                                                                                                       & \textbf{ResBlock}                                     \\ \cline{1-1} \cline{3-7} \cline{9-13} \cline{15-19} 
Channel and spatial attention residual{\cite{hu2019channel}}                            &  &                                                       & \cmark                                                    &                                                       &                                                       &                                                       &           &                                                       & \cmark                                                    &                                                       &                                                       &                                                       &           &                                                       & \cmark                                                    &                                                       &                                                       &                                                       \\
Enhanced Spatial Attention{\cite{liu2020residual}}                                        &  &                                                       &                                                       & \cmark                                                    &                                                       &                                                       &           &                                                       &                                                       & \cmark                                                    &                                                       &                                                       &           &                                                       &                                                       & \cmark                                                    &                                                       &                                                       \\
Two-Fold Attention Module{[}Ours{]}                                     &  &                                                       &                                                       &                                                       & \cmark                                                    & \cmark                                                    &           &                                                       &                                                       &                                                       & \cmark                                                    & \cmark                                                    &           &                                                       &                                                       &                                                       & \cmark                                                    & \cmark                                                    \\ \cline{1-1} \cline{3-7} \cline{9-13} \cline{15-19} 
Adaptive Residual Block{[}Ours{]}                                       &  &                                                       &                                                       &                                                       &                                                       & \cmark                                                    &           &                                                       &                                                       &                                                       &                                                       & \cmark                                                    &           &                                                       &                                                       &                                                       &                                                       & \cmark                                                    \\ \cline{1-1} \cline{3-7} \cline{9-13} \cline{15-19} 
\begin{tabular}[c]{@{}l@{}}PSNR on Set5 ($\times4$)\\ PSNR on Urban100 ($\times4$)\end{tabular} &  \rule{0pt}{2ex} \rule{0pt}{2ex}& 
\begin{tabular}[c]{@{}c@{}}$32.46$\\ $26.64$\end{tabular} & \rule{0pt}{2ex} \rule{0pt}{2ex}
\begin{tabular}[c]{@{}c@{}}$32.48$\\ $26.66$\end{tabular} & \rule{0pt}{2ex} \rule{0pt}{2ex}
\begin{tabular}[c]{@{}c@{}}$32.51$\\ $26.69$\end{tabular} & \rule{0pt}{2ex} \rule{0pt}{2ex}
\begin{tabular}[c]{@{}c@{}}$32.54$\\ $26.71$\end{tabular} & \rule{0pt}{2ex} \rule{0pt}{2ex}
\begin{tabular}[c]{@{}c@{}}$\boldsymbol{32.65}$\\ $\boldsymbol{26.79}$\end{tabular} & \rule{0pt}{2ex} \rule{0pt}{2ex}& \begin{tabular}[c]{@{}c@{}}$32.63$\\ $26.82$\end{tabular} & \rule{0pt}{2ex} \rule{0pt}{2ex}
\begin{tabular}[c]{@{}c@{}}$32.64$\\ $39.84$\end{tabular} & \rule{0pt}{2ex} \rule{0pt}{2ex}
\begin{tabular}[c]{@{}c@{}}$32.67$\\ $26.86$\end{tabular} & \rule{0pt}{2ex} \rule{0pt}{2ex}
\begin{tabular}[c]{@{}c@{}}$32.70$\\ $26.89$\end{tabular} & \rule{0pt}{2ex} \rule{0pt}{2ex}
\begin{tabular}[c]{@{}c@{}}$\boldsymbol{32.78}$\\ $\boldsymbol{26.96}$\end{tabular} & \rule{0pt}{2ex} \rule{0pt}{2ex}& \begin{tabular}[c]{@{}c@{}}$32.25$\\ $26.22$\end{tabular} & \rule{0pt}{2ex} \rule{0pt}{2ex}
\begin{tabular}[c]{@{}c@{}}$32.27$\\ $26.25$\end{tabular} & \rule{0pt}{2ex} \rule{0pt}{2ex}
\begin{tabular}[c]{@{}c@{}}$32.30$\\ $26.29$\end{tabular} & \rule{0pt}{2ex} \rule{0pt}{2ex}
\begin{tabular}[c]{@{}c@{}}$32.34$\\ $26.32$\end{tabular} & \rule{0pt}{2ex} \rule{0pt}{2ex}
\begin{tabular}[c]{@{}c@{}}$\boldsymbol{32.39}$\\ $\boldsymbol{26.41}$\end{tabular} \\ \hline
\end{tabular}}
\end{center}

\vspace{-25pt}
\end{table*}
\begin{table}[t!]
\caption{Impact of different Attention Mechanisms on MPRNet.}
% \vspace{-5pt}
\centering
\resizebox{\linewidth}{!}{
\begin{tabular}{|l|cccccc|}
\hline
% \multicolumn{6}{c}{Comparison of TFAM wrt other Attention Mechanisms on MPRNet} \\ 
% \hline 

\textbf{Dataset}&
\textbf{Baseline} &\rule{0pt}{2ex} \rule{0pt}{2ex} 
\textbf{SE}  & \rule{0pt}{2ex} \rule{0pt}{2ex}
\textbf{CBAM}  & \rule{0pt}{2ex} \rule{0pt}{2ex}
\textbf{CSAR} & \rule{0pt}{2ex} \rule{0pt}{2ex}
\textbf{ESA} & \rule{0pt}{2ex} \rule{0pt}{2ex}
\textbf{TFAM} \\ \hline
\small Set14 ($\times4$) &
$28.57$  &\rule{0pt}{2ex} \rule{0pt}{2ex}
$28.59$  &\rule{0pt}{2ex} \rule{0pt}{2ex}
$28.54$ &\rule{0pt}{2ex} \rule{0pt}{2ex}
$28.61$ &\rule{0pt}{2ex} \rule{0pt}{2ex}
$28.64$ &\rule{0pt}{2ex} \rule{0pt}{2ex}
$\textbf{28.67}$ \\
\small Urban100 ($\times4$) &
$26.19$ &\rule{0pt}{2ex} \rule{0pt}{2ex}
$26.21$  &\rule{0pt}{2ex} \rule{0pt}{2ex}
$26.18$ &\rule{0pt}{2ex} \rule{0pt}{2ex}
$26.23$ &\rule{0pt}{2ex} \rule{0pt}{2ex}
$26.25$ &\rule{0pt}{2ex} \rule{0pt}{2ex}
$\textbf{26.29}$ \\ \hline
\end{tabular}}

\label{tab:attention}
\vspace{-5pt}
\end{table}
\begin{table}[t!]
\caption{Effect of different configs of Residual Block and each learning pathway of the Adaptive Residual Block}
\vspace{-10.0pt}
\setlength\arrayrulewidth{0.5pt}
\begin{center}
\resizebox{\linewidth}{!}{
\begin{tabular}{|l|c|c|c|cccc|}
\hline 
% \multicolumn{8}{c}{Imapct of different ResBlock and each pathway of the ARB on proposed MPRNet}\\ \hline 

\textbf{Configs} & 
                   \begin{tabular}[c]{@{}c@{}}\textbf{MobileNet}\\ \textbf{BnBlock}\end{tabular} & 
                   \begin{tabular}[c]{@{}c@{}}\textbf{EDSR}\\ \textbf{ResBlock}\end{tabular}&
                   \begin{tabular}[c]{@{}c@{}}\textbf{RCAN}\\ \textbf{ResBlock}\end{tabular}&
                   \textbf{ARB$_{B}$} & \rule{0pt}{2ex}
                   \textbf{ARB$_{BA}$} & \rule{0pt}{2ex}
                   \textbf{ARB$_{R}$} & \rule{0pt}{2ex}
                   \textbf{ARB}   \\ \hline
     BN$_{p}$ &  &  &   & \cmark & \cmark  & \cmark & \cmark   \\ 
     Adp$_{p}$&  &  &   &    & \cmark  &  & \cmark    \\ 
     Res$_{p}$&  &  &   &  &   & \cmark & \cmark   \\ \hline\hline

    B100 ($\times4$)     & $27.24$   & $27.44$  & $27.52$ & $27.46$ & $27.58$ &  $27.55$ &  $\textbf{27.63}$  \\ 
    Urban100 ($\times4$) & $25.79$   &  $25.96$   & $26.08$ & $26.05$ & $26.15$ &  $26.11$ &  $\textbf{26.31}$  \\ \hline
\end{tabular}}
\vspace{-15pt}
\end{center}

\label{tab:Mblock}
\end{table}

% \small Set5            & $31.87$ & $31.91$ & $31.79$ & $32.21$ &  $32.17$ &  $\textbf{32.28}$  \\ 
%  \small Set14            & $28.23$ & $28.36$ & $27.95$ & $28.52$ &  $28.44$ &  $\textbf{28.62}$  \\ 
% Please add the following required packages to your document preamble:
% \usepackage{multirow}
\begin{table}[]
\caption{Study on combining different Residual Connections.}
% \vspace{-5pt}
\centering
\footnotesize
\resizebox{\linewidth}{!}{
\begin{tabular}{lccccccc|}
% \toprule
\hline
% \multicolumn{8}{c}{Impact of different Residual Learning Connections on proposed MPRNet}                                                                                                                                                                                                                                                                                                                                                                                       \\ \hline
\multicolumn{2}{|l|}{\textbf{Options}}                                                                             & \textbf{Baseline}                                              & $\boldsymbol{1st}$                                                   & $\boldsymbol{2nd}$                                                   & $\boldsymbol{3rd}$                                                   & $\boldsymbol{4th}$                                                   & \multicolumn{1}{c|}{$\boldsymbol{5th}$}                                                   \\ \hline
\multicolumn{1}{|l|}{\multirow{3}{*}{Residual Learning Connections}} & \multicolumn{1}{c|}{\textbf{LRC}}  & \xmark                                                   & \cmark                                                   &                                                       &                                                    &  \cmark                                                     & \multicolumn{1}{c|}{\cmark}                                                   \\ \cline{2-2}
\multicolumn{1}{|l|}{}                                               & \multicolumn{1}{c|}{\textbf{GRC}}  & \xmark  &   & \cmark &   & \cmark                                                       & \multicolumn{1}{c|}{\cmark}                                                   \\ \cline{2-2}
\multicolumn{1}{|l|}{}                                               & \multicolumn{1}{c|}{\textbf{LRSC}} & \xmark  &   &  & \cmark &                                                     & \multicolumn{1}{c|}{\cmark}                                                   \\ \hline
\multicolumn{2}{|l|}{\begin{tabular}[c]{@{}l@{}}PSNR on Set5 ($\times3$)\\ PSNR on Urban100 ($\times3$)\end{tabular}}  & 
\begin{tabular}[c]{@{}c@{}}$34.42$\\ $28.30$\end{tabular} & 
\begin{tabular}[c]{@{}c@{}}$34.40$\\ $28.29$\end{tabular} & \rule{0pt}{2ex} \rule{0pt}{2ex} \rule{0pt}{2ex}
\begin{tabular}[c]{@{}c@{}}$34.47$\\ $28.35$\end{tabular} & \rule{0pt}{2ex} \rule{0pt}{2ex} \rule{0pt}{2ex}
\begin{tabular}[c]{@{}c@{}}$34.45$\\ $28.33$\end{tabular} & \rule{0pt}{2ex} \rule{0pt}{2ex} \rule{0pt}{2ex}
\begin{tabular}[c]{@{}c@{}}$34.52$\\ $28.38$\end{tabular} & \rule{0pt}{2ex} \rule{0pt}{2ex} \rule{0pt}{2ex} \begin{tabular}[c]{@{}c@{}}$\boldsymbol{34.57}$\\$\boldsymbol{28.42}$\end{tabular} \\ \hline
\end{tabular}}
\label{tab:RLC}
\vspace{-16pt}
\end{table}
\subsection{Ablation Study}\label{sec:study}
To further investigate the performance of the proposed model, a deep analysis on the Two-Fold Attention Module, the Adaptive Residual Block, and Residual Learning Connections is performed via an extensive ablation study.

%Firstly, the effect of each components in Adaptive Residual Block on the final performance of the proposed model is investigated. Secondly, the proposed Two-Fold Attention Mechanism is studied and its performance is compared with respect to other attention mechanisms.
\textsc{\bf Two-Fold Attention Module}\label{sec:CA}. 
In this section, Deep investigation of the impacts of our proposed TFAM on SOTA SR models are provided. The performance of image SR has improved greatly with the application of Attention Mechanism (AM). Table \ref{tab:channel_com} shows the performance of applying recent AMs including Channel and spatial attention residual (CSAR) \cite{hu2019channel}, Enhanced Spatial Attention (ESA)\cite{liu2020residual}, and our Two-Fold Attention Module (TFAM) on EDSR, RCAN, and MSRN. For a fair comparison, all the models were re-trained with their default setting and AMs are added to the end of their Block, and replaced in the same place as RCAN's Channel Attention placed. As can be seen, by using the aforementioned attention module, the performance of the baseline models are increased that shows the importance of AM in SR tasks. By applying the CSAR to the mentioned approaches, PSNR improves in EDSR and MSRN but does not show enough improvement in RCAN. In contrast, ESA is enhanced version of CASR, which combine both the channel and spatial information, improves all the baseline models. However ESA cannot completely boost the power of the networks due to lack of highlighting informative feature in spatial information. For this propose, we introduce Two-Fold Attention Module, which consider both channel and spatial information and maximize the performance of the networks. TFAM extracts the channel and spatial statistic among channels and spatial axis to further enhance the discriminative ability of the network. As a results, TFAM shows better performance than all the aforementioned ones and boosted the baseline SOTA.

Furthermore, Table \ref{tab:attention} contains the study on impact of recent AMs on our MPRNet. Namely, SE\cite{hu2018squeeze}, CBAM \cite{woo2018cbam}, CSAR \cite{hu2019channel}, ESA \cite{liu2020residual}, and TFAM. We apply all the aforementioned AM to our ARB blocks and Feature Module, and provide the performance. The proposed MPRNet with CBAM, could not achieve better results than baseline or SE due to losing channel information and applying the Max-pooling in CA unit which shows harm the performance. Unlike, MPRNet with CASR achieves better results than CBAM and SE because of considering both channel and spatial information but not better than ESA. However, our TFAM performs better among all the AMs by calculating the first order statistics on CA unit and applying Avg- and Max-pooling operations along the channel axis, which is effective in highlighting informative regions and extracts the most important features like edges.

Table \ref{tab:channel_com} also shows the efficiency of our ARB with conjunction of TFAM when it is applied to other SOTA models. As indicated, ARB with TFAM together can improve the PSNR of SOTA models with a large margin.

% , which shows that proposed ARB can also boost the performance of TFAM
% Table \ref{tab:pooling} shows the effect of different pooling layers inside the proposed TFAM on the reconstruction results. Three different pooling layers are considered for the CA unit (Average, Max, and Power-Average pooling), while two pooling layers are considered for the Pos unit (Average and Max pooling); each pooling layer is in charge of extracting different information. The study firstly considers the different possibilities of pooling layers in the CA unit and find the best pooling operation then combines with Pos unit. The best result is achieved by using Global Average pooling in the CA unit and concatenating Average and Max pooling in the Pos unit.  

\textsc{\bf Adaptive Residual Block}\label{sec:lmbAS}. Table \ref{tab:Mblock} presents the impact of different Residual Blocks and the proposed Adaptive Residual Block (ARB) on our MPRNet. In this work, three different structures of residual blocks from SOTA models are considered to compare with our proposed ARB, namely, MobileNet-BottleneckBlock, EDSR-ResBlock, RCAN-ResidualChannelBlock. All the models were trained with the same settings. As can be seen, MobileNet-BottleneckBlock could not perform well in SR tasks due to difficulty of extracting high-frequency information and gradient confusion. EDSR-ResBlock is the ResNet without batch normalization layer, but still could not achieve good results due to the lack of extracting rich feature maps and eliminating noises from LR feature space. RCAN-ResidualChannelBlock performs better than aforementioned ResBlock due to channel attention in their structure. However RCAN-ResidualChannelBlock did not show better results than our proposed ARB since our ARB can learn more expressive spatial information, have access to high-dimensional information and also with the help of TFAM can maximize the whole performance of block.

% a deep study on combinations of different learning pathways in our proposed ARB is provided.
Additionally, effect of each learning pathways of ARB on the performance is provided.
$ARB_{B}$, $ARB_{BA}$ and $ARB_{R}$ are Adaptive Residual Block with bottleneck path; ARB with bottleneck and adaptive paths; ARB with bottleneck and residual path respectively. As shown in Table \ref{tab:Mblock}, MPRNet with all learning pathways (ARB) achieves the best performance among all the mentioned ResBlock and combinations of different ARB learning pathways. This is caused by allowing the main parts of network to focus on more informative components of the LR features and force the network to focus more on abstract features, which are important in SR tasks. Furthermore, the proposed pathways helps the model to converge better and performs better than all the baseline models. In a nutshell, information propagates locally via residual path, adaptively extract the informative features via adaptive path, and learn more meaningful spatial information by Bottleneck path. By doing so, information is transmitted by multiple pathways inside of ARB and main parts of network access to more expressive and richer feature maps, resulting in superior PSNR.

%  \cred{On the grounds that ARB proposed to extract informative features from LR space by accumulating all the components together.}
\textsc{\bf Effect of Residual Learning Connections.} Table \ref{tab:RLC} shows the extensive study of the impact of Residual Learning Connections on our design of MPRNet, i.e. Local Residual Connection (LRC), Global Residual Connection (GRC), and Long Range Skip Connection (LRSC). In this work, residual connections except LRSC comprise concatenation followed by a 1$\times$1 Conv layer. As we can see, MPRNet without any residual connection performs relatively low (i.e. baseline). However, MPRNet with only GRC in Residual Module shows better performance than baseline since GRC transports the information from mid- to high-layers and helps the model to better leverage multi-level representations by collecting all information before the next module. 

On the contrary, MPRNet with only LRC inside Residual Concatenation Block could not perform better than the MPRNet with GRC. This behavior was expected as mentioned in \cite{he2016identity} that 1$\times$1 Conv layer on the residual connection can confuse optimization and prevent information propagation due to multiplicative manipulations. However, MPRNet can show better performance by using both connections ($4th$ col.). This is due to GRC eases the information propagation issue that LRC suffers from.

To end this, LRSC also added to the MPRNet to carry the shallow information to high-level layers. Thus, information is transferred by multiple connections, which mitigates the vanishing gradient problem and network has access to multi-level representation. As a results, MPRNet with all connections ($5th$ col.) can performs greatly better.    

% \textsc{\bf Effect of Optimizer.} Table \ref{tab:optim} shows the investigation on the effect of the well-known Adam and AdamP\cite{heo2020slowing} optimizer on the reconstruction results. We discovered that AdamP constantly perform better than Adam on all benchmark datasets for all scale factors with a large margin. 

\textsc{\bf Model Complexity Analysis.} Figure \ref{fig:ParamsMac} indicates the comparison regard to the model size and PSNR with $15$ recent state-of-the-art SR models. Our MPRNet achieves the best performance among all the lightweight SR approaches with much fewer parameters and achieves better or comparable results when compared with computationally expansive models. This shows that our MPRNet is well-balanced in terms of model size and reconstruction results.

% . Compared

% with DBPN, our RFANet achieves much higher PSNR with
% a slightly larger model
%------------------------------------------------------------------------
\vspace{-5pt}
\section{Conclusions}
This paper proposes a novel lightweight network (MPRNet) that achieves the best performance against all existing lightweight SOTA approaches. The main idea behind of this work is to design an advanced lightweight network to deliver almost similar results to heavy computational networks. %To achieve this, and focus on learning high-level information, 
A novel Residual Module is proposed to let abundant low-level information to be avoided through multiple connections. In addition, an efficient Adaptive Residual Block is proposed to allows MPRNet achieves more rich feature-maps through the multi-path learning. Furthermore, to maximize the power of the network a Two-Fold Attention Module is proposed, which refine the extracted information along channel and spatial axes to further enhance the discriminative ability of the network. Extensive evaluations and comparisons are provided.

%on BI, BD, and DN degradation models well demonstrate the effectiveness of our MPRNet in terms of both quantitative and qualitative results while retaining fairly low computation and memory requirements.

\section*{Acknowledgment}
This work has been partially supported by the Spanish Government under Project TIN2017-89723-P; the ``CERCA Programme / Generalitat de Catalunya"; and the ESPOL project CIDIS-56-2020.

% Both qualitative and quantitative results show that the proposed Two-Fold Attention Mechanism together with the proposed Adaptive Residual Block can outperform all the lightweight approaches and achieves competitive results against expensive computational networks.  

%-------------------------------------------------------------------------

\balance
{\small
\bibliographystyle{ieee_fullname}
\bibliography{egbib}
}

\end{document}